\documentclass[journal]{IEEEtran}
\usepackage[T1]{fontenc}
\usepackage{textcomp}
\usepackage[latin9]{inputenc}
\usepackage{array}
\usepackage{algorithm}
\usepackage{algpseudocode}
\usepackage{float}
\usepackage{xcolor}
\usepackage{multirow}
\usepackage{amsmath}

\usepackage{tabularx}
\usepackage{array}

\newcolumntype{C}[1]{>{\centering\arraybackslash}m{#1}}
\newcolumntype{Y}{>{\centering\arraybackslash}X}

\usepackage{amssymb}
\usepackage{graphicx}
\usepackage[bookmarks=true,bookmarksnumbered=true,bookmarksopen=true,bookmarksopenlevel=1,
breaklinks=false,pdfborder={0 0 0},pdfborderstyle={},backref=false,colorlinks=false]
{hyperref}

\hypersetup{
	pdftitle={Multiport Analytical Pixel Electromagnetic Simulator (MAPES) for AI-assisted RFIC and Microwave Circuit Design},
	pdfauthor={Junhui Rao et al.},
	pdfpagelayout=OneColumn, pdfnewwindow=true, pdfstartview=XYZ, plainpages=false
}

\makeatletter

\let\oldforeign@language\foreign@language
\DeclareRobustCommand{\foreign@language}[1]{%
	\lowercase{\oldforeign@language{#1}}}

\usepackage[caption=false,font=footnotesize]{subfig}

\makeatother
\begin{document}
	
	\bstctlcite{IEEEexample:BSTcontrol}
	\title{Multiport Analytical Pixel Electromagnetic Simulator (MAPES) for AI-assisted
		Microwave Circuit and RFIC Design}
	\author{Junhui~Rao,~\IEEEmembership{Member,~IEEE,} Yi~Liu,~\IEEEmembership{Member,~IEEE,}  Jichen Zhang, \IEEEmembership{Student Member, IEEE,} Zhaoyang Ming, \IEEEmembership{Student Member, IEEE,} Tianrui~Qiao,~\IEEEmembership{Member,~IEEE,}  Yujie Zhang ,~\IEEEmembership{Member,~IEEE,} Chi Yuk~Chiu,~\IEEEmembership{Senior Member,~IEEE,} ~Hua~Wang,~\IEEEmembership{Fellow,~IEEE,} and ~Ross~Murch,~\IEEEmembership{Fellow,~IEEE}\thanks{This work was supported by Hong Kong Research Grants Council Area of Excellence Grant AoE/E
			601/22-R. (\textit{Corresponding author: Junhui Rao.})}\thanks{Junhui Rao, Yi~Liu,  Jichen Zhang, Zhaoyang Ming, Tianrui~Qiao and Chi Yuk Chiu are with the Department of Electronic and Computer Engineering, the Hong Kong University of Science and Technology, Hong Kong (e-mail: \protect\href{mailto:jraoaa@connect.ust.hk}{jraoaa@connect.ust.hk}).}
		\thanks{ Yujie Zhang is with School of Electrical and Electronic Engineering, Nanyang Technological University (NTU), Singapore (e-mail:yujie.zhang@ntu.edu.sg).}
		\thanks{Hua Wang is with the School of Information Technology and Electrical Engineering, ETH Z\"urich, 8092 Z\"urich, Switzerland (e-mail:huawang@ethz.ch).}
		\thanks{R. Murch is with the Department of Electronic and Computer Engineering and Institute for Advanced Study (IAS) at the Hong Kong University of Science and Technology, Hong Kong    (e-mail: \protect\href{mailto:eermurch@ust.hk}{eermurch@ust.hk}).}}

	\markboth{}{Junhui Rao \MakeLowercase{\emph{et al.}}: babba}
	\maketitle
	\begin{abstract}
		This paper proposes a novel analytical framework, denoted the Multiport Analytical Pixel Electromagnetic Simulator (MAPES). MAPES enables efficient and accurate prediction of the electromagnetic (EM) performance of arbitrary pixel-based microwave (MW) and RFIC structures. \textcolor{black}{Unlike the Internal Multiport Method (IMPM), which optimizes only connecting elements within a fixed, gap-separated pixel skeleton, MAPES operates directly on the all-pixel presence/absence formulation used in recent MW/RFIC design. This is enabled by diagonal virtual pixels, an occupancy-to-load mapping, and a multi-layer/via port-level formulation that have no counterpart in IMPM.} By introducing virtual pixels and diagonal virtual pixels and inserting virtual ports at critical positions, MAPES captures all horizontal, vertical, and diagonal electromagnetic couplings within a single multiport impedance matrix. Only a small set of full-wave simulations (typically about 1\% of the datasets required by AI-assisted EM emulators) is needed to construct this matrix. Subsequently, any arbitrary pixel configuration can be evaluated analytically using a closed-form multiport relation without additional full-wave calculations. The proposed approach eliminates data-driven overfitting and ensures accurate results across all design variations. Using MAPES, comprehensive examples for single- and double-layer PCBs and CMOS processes (180 nm and 65 nm) confirm that high prediction accuracy with 600-2000$\times$ speed improvement is achieved compared to CST simulations. Owing to its efficiency, scalability, and reliability, MAPES provides a practical and versatile tool for AI-assisted MW circuit and RFIC design across diverse fabrication technologies.
	\end{abstract}
	
	
\section{Introduction}

\IEEEPARstart{A}{s} 5G is deployed and 6G research advances~\cite{5G1,5G2,6G1,6G2}, an
increasing variety of microwave (MW) circuits and RFICs, such as matching networks,
combiners, splitters, and couplers, must be designed under diverse specifications.
\textcolor{black}{Conventional design relies heavily on designer experience and a small set
	of tunable geometric parameters (e.g., transmission-line widths and lengths), which
	restricts the searchable design space and the achievable
	performance~\cite{Karahan2023princeton,Chen2011tra,Lin2005tra,Wilkinson1960tra,Yue1998tra}.}

\textcolor{black}{To enlarge the design space, AI-assisted methods discretize the layout
	into a pixel array and treat the binary presence/absence of each pixel as the design
	variable~\cite{Chu2025ETH,Munzer2020,Chu2025eth1,Guo2025princeton,Karahan2023princeton,Karahan2024princeton,Chen2026TAP,Chenna2026,Chenna2026a}.
	Since full-wave evaluation of so many candidates is infeasible, CNN-/deep-CNN-based EM
	emulators are trained on large full-wave datasets to predict S-parameters, after which
	inverse synthesis is performed using heuristic or generative
	algorithms~\cite{Karahan2023princeton,Karahan2024princeton_NC,Gupta2023princeton,Guo2025princeton}.}
\textcolor{black}{However, these emulators inherit three well-known drawbacks of data-driven
	models: the training datasets are costly to generate (e.g., 180k samples for a
	$25\times25$ grid require $\sim$13 days on 400 CPU cores~\cite{Karahan2024princeton_NC}),
	they cover only a negligible fraction of the design space and are thus prone to overfitting
	on out-of-distribution layouts~\cite{Goodfellow2016,Srivastava2014}, and they impose
	substantial storage/I/O overhead.}

\textcolor{black}{The all-pixel presence/absence formulation used in AI-assisted methods is the key source of design
	freedom  in ongoing MW and RFIC design. Treating every pixel as an independent binary
	degree of freedom increases the searchable space exponentially with the number of pixels
	(e.g., $2^{256}$ layouts for a $16\times16$ grid). Consequently, arbitrary and non-intuitive metal
	shapes, including isolated islands and diagonal-only links, can emerge to realize EM
	responses that are unreachable by conventional or connecting-element layouts, which is
	precisely why pixel-based AI-assisted MW/RFIC design has grown rapidly in recent years.}

\textcolor{black}{On the other hand, a separate line of research originates from
	pixel-antenna/surface
	design~\cite{Araque2006,Song2014,Shen2018,Zhang2024,Rao2025,Rao2025a,Rao2025b,Zhang2025},
	where the structure is discretized into gap-separated pixels and the \emph{connecting elements} between adjacent fixed pixels are optimized. For this formulation, multiport
	methods such as the Internal Multiport Method (IMPM)~\cite{Song2014,Shen2018}
	provide highly efficient and accurate performance evaluation by inserting ports at the
	inter-pixel gaps. However, IMPM is intrinsically restricted to optimizing the connecting elements within a fixed grid and cannot operate on the presence/absence of all pixels.  It always requires a fixed skeleton of permanently present metal.}

\begin{figure}[t]
	\centering
	\includegraphics[width=0.95\columnwidth]{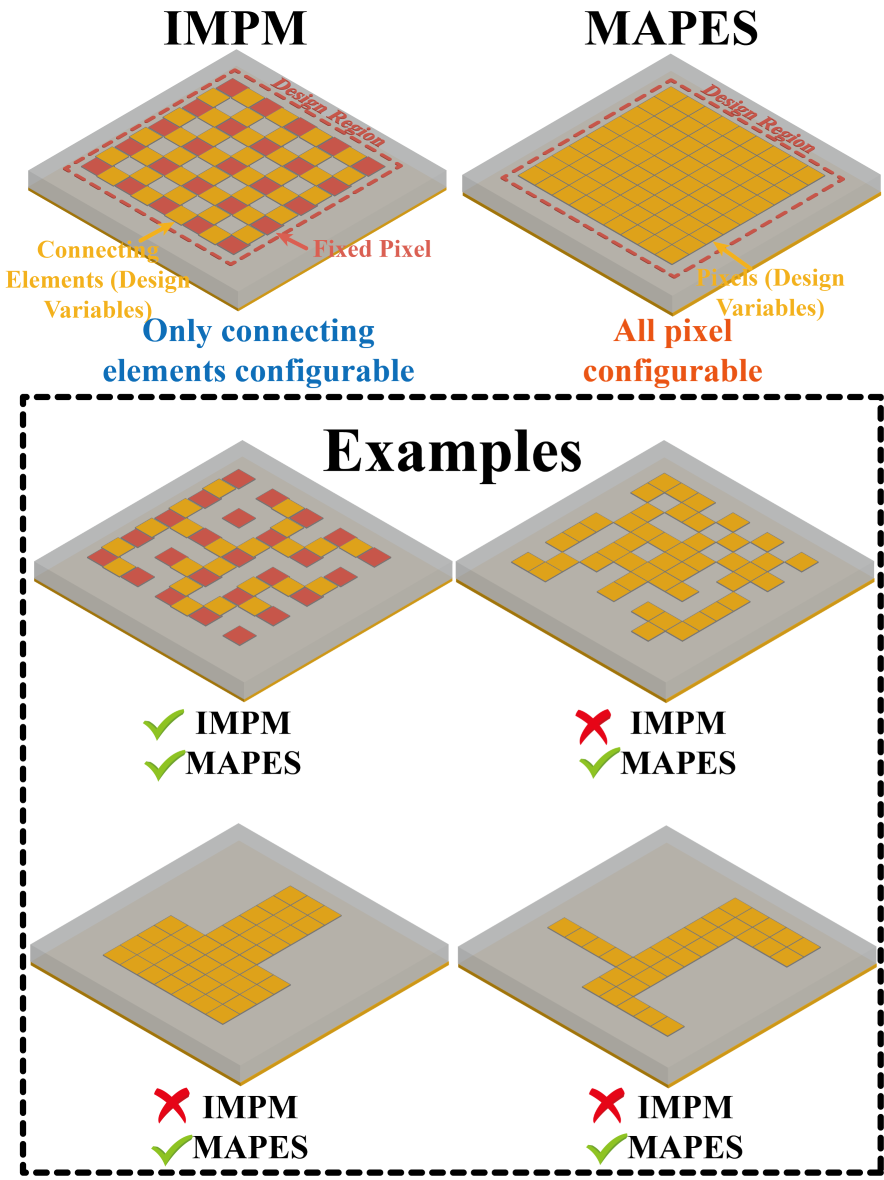}
	\caption{\textcolor{black}{Comparison of the design domains of IMPM and the proposed
			MAPES. The example layouts illustrate that contiguous
			presence/absence patterns, including those relying on diagonal-only connections, can
			be represented by MAPES but not by IMPM.}}
	\label{fig:impm_vs_mapes}
\end{figure}

	\textcolor{black}{The two techniques above have incompatible design variables. AI-assisted design uses the binary presence/absence of \emph{contiguous} pixels, whereas IMPM assumes permanently present, gap-separated pixels and optimizes only the connecting elements between them (Fig.~\ref{fig:impm_vs_mapes}). Therefore, IMPM cannot be directly used for the all-pixel presence/absence formulation for three reasons: (i)~its inputs are connecting elements rather than pixel occupancy states; (ii)~it does not capture diagonal coupling between contiguous pixels, which is essential in RFICs and cannot be introduced through the extra overlapping layers/vias used in pixel antennas~\cite{Jing2022}; and (iii)~its I/O ports have fixed positions and numbers, whereas RFIC/MW synthesis requires arbitrary I/O placement. Bridging this gap would combine the efficiency of multiport theory with the full design flexibility of pixel-based MW/RFIC synthesis.}

\begin{table}[t]
	\textcolor{black}{
		\caption{Comparison Between IMPM and the Proposed MAPES}
		\label{tab:impm_vs_mapes}
		\centering
		\renewcommand{\arraystretch}{1.3}
		\setlength{\tabcolsep}{3pt}
		\footnotesize
		\newcommand{\yes}{\textcolor{green!55!black}{\checkmark}}
		\newcommand{\no}{\textcolor{black}{\boldmath$\times$}}
		\begin{tabular}{
				>{\raggedright\arraybackslash}m{0.24\columnwidth}
				>{\centering\arraybackslash}m{0.30\columnwidth}
				>{\centering\arraybackslash}m{0.32\columnwidth}
			}
			\hline
			\textbf{Aspect} & \textbf{IMPM~\cite{Song2014,Shen2018}} & \textbf{MAPES (This Work)} \\
			\hline
			Design variable & Connecting elements only & \textbf{\yes\ All pixels (presence/absence)} \\
			Configurable design region & \no\ Partial (fixed + connecting elements) & \textbf{\yes\ Full design region} \\
			Pixel layout & Separated by gaps & \textbf{\yes\ Contiguous, no gaps} \\
			Diagonal EM coupling & \no\ Not captured & \textbf{\yes\ Captured (diagonal virtual pixels)} \\
			I/O ports & Fixed position \& number & \textbf{\yes\ Arbitrary position \& number} \\
			Target application & Pixel antennas \& surfaces & \textbf{MW circuits \& RFICs} \\
			\hline
		\end{tabular}
		\vspace{2pt}\\
		\begin{minipage}{0.96\columnwidth}
			\scriptsize MAPES reformulates the design domain from connecting elements to the presence/absence of all pixels, and introduces the constructs (diagonal virtual pixels, occupancy-to-load mapping, and a multi-layer/via and arbitrary-I/O port formulation) that are required for, and have no counterpart in, IMPM.
		\end{minipage}
	}
\end{table}

\begin{figure}[t]
	\centering
	\includegraphics[width=0.95\columnwidth]{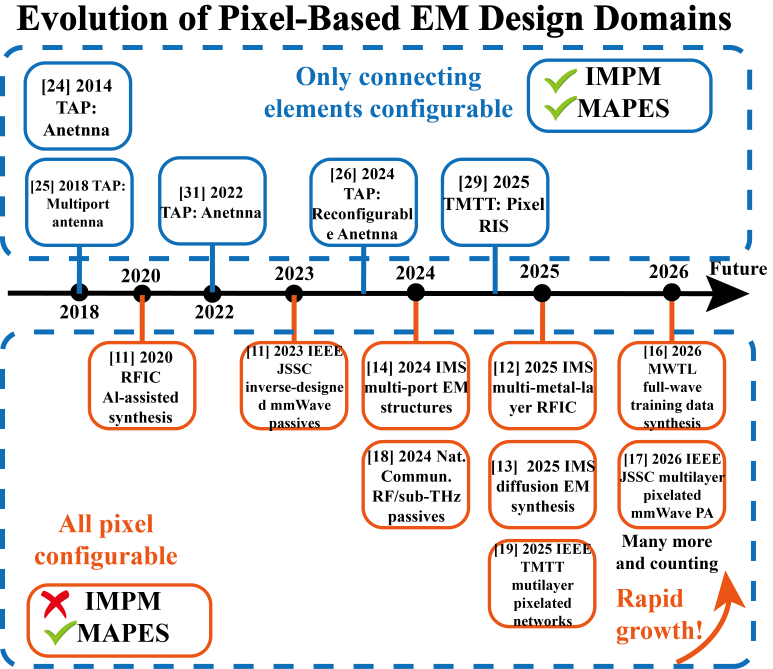}
	\caption{\textcolor{black}{Evolution of pixel-based EM design domains. Earlier
			pixel-antenna works adopt the ``only connecting elements configurable'' formulation
			(e.g., IMPM), whereas recent AI-assisted MW-circuit/RFIC works adopt the ``all pixels
			configurable'' (presence/absence) formulation, which has grown rapidly in recent years.
			The proposed MAPES is compatible with both formulations, bridging multiport analysis
			and AI-assisted pixel-based design.}}
	\label{fig:pixel trending}
\end{figure}

\textcolor{black}{In this paper we propose the Multiport Analytical Pixel Electromagnetic
	Simulator (MAPES), which bridges this gap by reformulating pixel presence/absence states
	as the design variables within a multiport analytical framework
	(Fig.~\ref{fig:impm_vs_mapes}) through three innovations with no counterpart in IMPM:
	(i)~diagonal virtual pixels that capture all horizontal, vertical, and diagonal coupling
	paths; (ii)~a mapping algorithm that converts any binary occupancy matrix into a load
	matrix; and (iii)~a port-level formulation supporting multi-layer structures with vias and
	arbitrary I/O placement. All EM characteristics of the design space are condensed into a
	single impedance matrix, so any pixel configuration is evaluated through one closed-form
	multiport relation. As summarized in Fig.~\ref{fig:impm_vs_mapes} and
	Table~\ref{tab:impm_vs_mapes}, MAPES is not a straightforward modification of IMPM but differs fundamentally in both formulation and capability, extending analytical multiport modeling to operate
	directly on the full-pixel presence/absence layout. The
	main contributions are summarized as follows:}

\begin{enumerate}
	
	\item \textcolor{black}{\textit{A new presence/absence multiport formulation:}
		Departing from connection-based multiport methods, MAPES introduces diagonal virtual
		pixels and virtual ports so that any contiguous pixel pattern is converted into a load
		matrix and evaluated analytically by multiport network theory, capturing all horizontal,
		vertical, and diagonal couplings that connection-based methods such as IMPM cannot
		represent.}

	\item \textcolor{black}{\textit{Generalization without overfitting:} predictions follow
		rigorously from microwave theory rather than data-driven regression, ensuring accuracy
		for arbitrary pixel patterns and I/O ports.}

		\item \textcolor{black}{\textit{Small prior dataset:} MAPES requires only a one-time set of
			full-wave simulations to build $\mathbf{Z}_{\mathrm{ALL}}$, typically about 1\% of the
			training data of AI-assisted emulators (e.g., 3.5K vs.\ 180K for a $25\times25$
			grid~\cite{Karahan2024princeton_NC}).}
	
	\item \textcolor{black}{\textit{Process generality:} the small prior dataset and low cost
		let MAPES scale to multi-layer PCBs and various RFIC processes, including vias between
		layers.}
	
	\item \textcolor{black}{\textit{Experimental verification:} single-/double-layer PCB and
		CMOS 180\,nm/65\,nm examples show good agreement with full-wave results and a
		$600\text{--}2000\times$ speedup.}
	
\end{enumerate}		

\textcolor{black}{It should be noted that the scope of MAPES is restricted to planar, layered
	microwave and RFIC structures whose layout can be discretized into a contiguous pixel array
	over one or more metal layers, with optional inter-layer vias. Within this domain, MAPES
	evaluates \emph{arbitrary pixel patterns}, i.e., the presence/absence of every pixel across
	the combinatorially large design space.
	MAPES is therefore not intended for arbitrary three-dimensional electromagnetic structures
	that cannot be represented by such a pixelated, layered formulation; rather, its generality
	lies in handling any pixel configuration within this design domain through a single closed-form
	multiport relation.}
	
		A related class of methods precomputes an impedance representation for a ``mother''
	structure and reuses it across layout variations~\cite{Johnson1999}; a detailed comparison
	with MAPES, including PNGF-style methods~\cite{Sun2025}, is given in Section~IV-D.

	The remainder of this paper is organized as follows. Section~II details the MAPES
	methodology. Section~III validates MAPES through four case studies spanning single-/double-
	layer PCB and CMOS 180\,nm/65\,nm structures, with S-parameter comparisons against CST.
	Section~IV compares MAPES with full-wave solvers, AI-assisted emulators, and
	precomputation-based calculators, and discusses the modeling assumptions, scalability, and
	physical properties. Section~V concludes the paper.
			
		\begin{figure*}[t]
			\begin{centering}
				\textsf{\includegraphics[width=2\columnwidth]{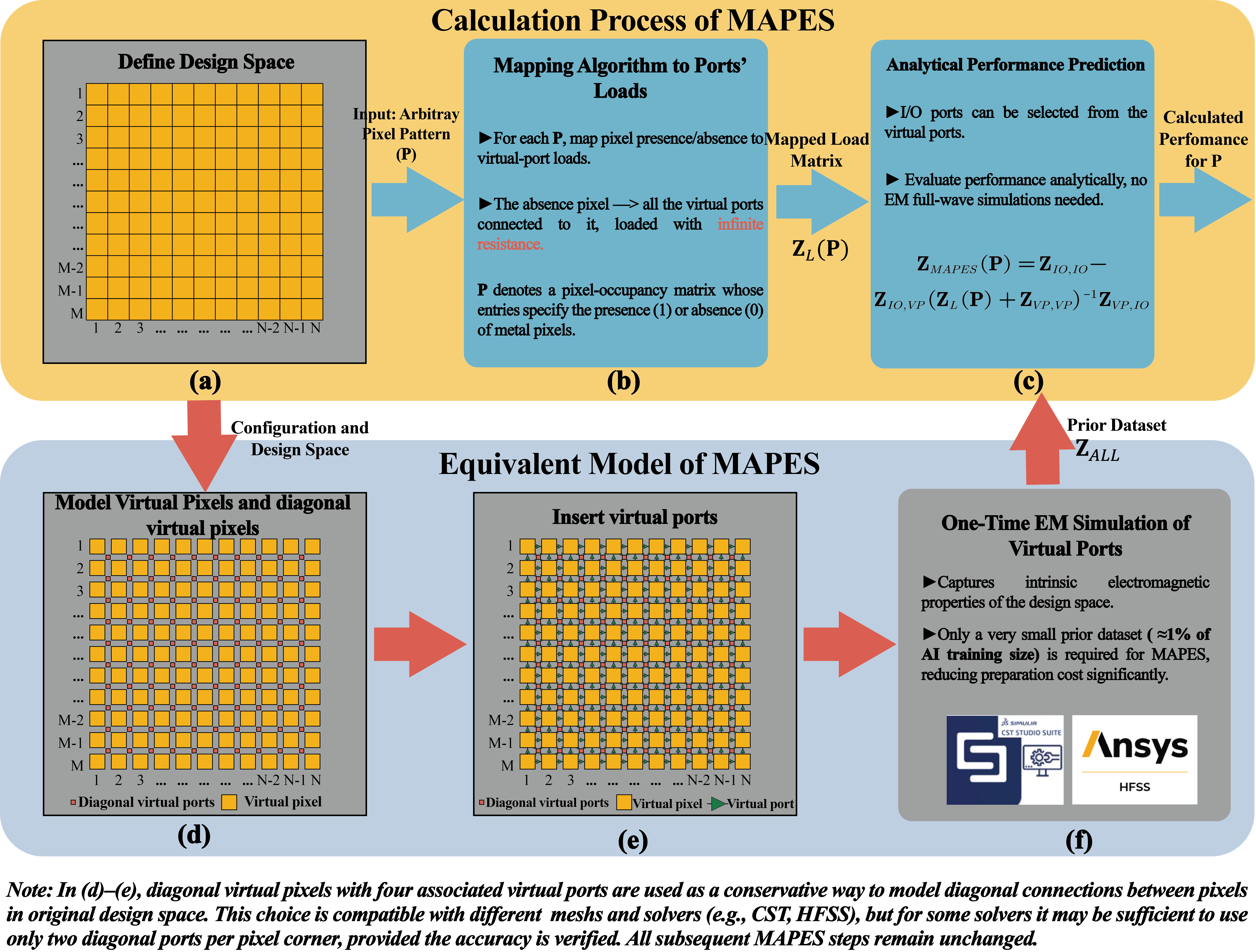}}
				\par\end{centering}
			
			\caption{MAPES workflow. (a) Pixel occupancy matrix \(\mathbf{P}\); (b) map \(\mathbf{P}\) to the virtual-port load matrix \(\mathbf{Z}_L(\mathbf{P})\); (c) analytically compute the I/O response via the closed-form multiport relation; (d) construct virtual pixels with diagonal virtual pixels; (e) insert virtual ports; (f) one-time full-wave extraction of \(\mathbf{Z}_{\mathrm{ALL}}\).}
			
			\label{working_flow}
		\end{figure*}

	\section{Methodology of MAPES}
	
	The core concept of MAPES is illustrated in Fig.~\ref{working_flow} where the calculation process and the equivalent model are shown. Starting from the original physical pixel design space in Fig.~\ref{working_flow}(a), we construct an equivalent model based on a modified virtual pixel structure, as shown in Fig.~\ref{working_flow}(d). Virtual ports are inserted between the virtual pixels (Fig.~\ref{working_flow}(e)) and these virtual ports can be loaded with connections to form any equivalent structure in the design space (Fig.~\ref{working_flow}(a)). This virtual model is analyzed using efficient techniques (described later) to accurately predict the performance of the original design space, as illustrated in Fig.~\ref{working_flow}(c). The construction and use of this equivalent model proceeds through several well-defined steps.
	
	We first create an array of virtual pixels and insert diagonal virtual pixels at the pixel corners, as in Fig.~\ref{working_flow}(d). Virtual ports are then introduced to capture all relevant horizontal, vertical, diagonal, and (optionally) inter-layer interactions across the design space (Fig.~\ref{working_flow}(e)). A single round of full-wave simulations for these virtual ports yields the impedance matrix $\mathbf{Z}_{\mathrm{ALL}}$, which embeds the intrinsic EM couplings of the entire design space. This matrix is computed only once, greatly reducing the overall computational burden of MAPES. We denote the "design space", $\mathbf{P}$, as a pixel-occupancy matrix whose entries specify the presence ($1$) or absence ($0$) of metal pixels (or vias) at each location in the discretized MW/RFIC layout domain in Fig.~\ref{working_flow}(a). For any physical pixel configuration in the original design space (Fig.~\ref{working_flow}(a)), a straightforward mapping algorithm converts the presence/absence pattern ($\mathbf{P}$) into a load matrix on the non-I/O virtual ports (Fig.~\ref{working_flow}(b)). Using a closed-form multiport method, MAPES then computes the I/O response $\mathbf{Z}_{\mathrm{MAPES}}(\mathbf{P})$ directly from $\mathbf{Z}_{\mathrm{ALL}}$ and the mapped loads, without any additional full-wave simulations. This enables fast and accurate performance prediction and efficient exploration of large MW/RFIC pixel design spaces. 
	
	The details of these steps are presented in Subsection II A to D. Throughout this paper, the terms "pixel" and "pixel array" refer to the pixels in the original design space, in which pixels are placed contiguously without gaps. In contrast, "virtual pixel" and "diagonal virtual pixel" denote the elements of the equivalent model used in MAPES.
	
	\subsection{Virtual Pixel Configurations for MAPES}
	To characterize the physical properties across the entire design space, a comprehensive equivalent model needs to be developed. In this work we utilize an efficient approach based on virtual pixels and virtual ports. Because the original design space is formed by an array of seamlessly placed pixels, direct insertion of virtual ports is not feasible. Therefore, we first construct an alternative virtual pixel array over the same design area and augment it with diagonal virtual pixels, as shown in  Fig. \ref{working_flow}(d). 
	
\begin{figure}[t]
		\begin{centering}
			\textsf{\includegraphics[width=1\columnwidth]{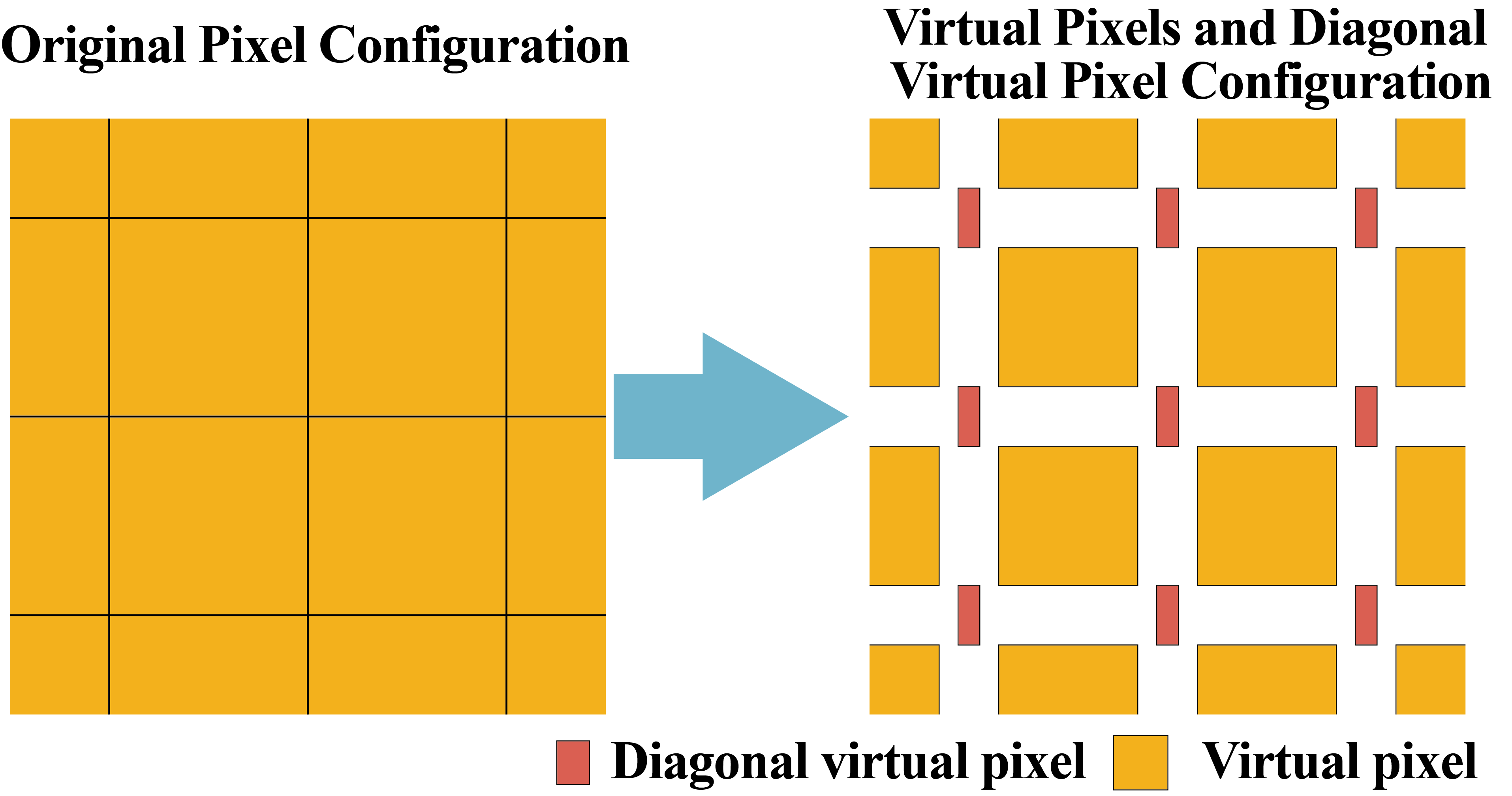}}
			\par\end{centering}
		\caption{Magnified view of the design space (left) and the equivalent model (right). The model (right) consists of virtual pixels and diagonal virtual pixels in  MAPES.}
		\label{step2}
	\end{figure}

	First, the original pixels are mapped to virtual pixels in a one-to-one manner, maintaining their central locations while slightly reducing their physical dimensions to allow for the insertion of virtual ports between them. However, this configuration alone does not fully capture the relationships between diagonally connected pixels. To address this, we introduce an array of diagonal virtual pixels located at the corners of the virtual pixels. These diagonal virtual pixels are smaller than the gaps between the virtual pixels, as shown in the detailed view  of Fig. \ref{step2}.
	
	It should be noted that the use of diagonal virtual pixels with four virtual ports connected to each of them (as introduced in the following subsection), is a conservative modeling choice intended to accurately capture diagonal connections between pixels in the original design space. This construction is compatible with various mesh settings (e.g., hexahedral and tetrahedral meshes) and full-wave solvers (e.g., the CST transient solver and the HFSS frequency-domain solver). For some solvers, however, using diagonal virtual pixels may be unnecessary, and the scheme can be simplified to using only two diagonal ports at each pixel corner, provided that the resulting accuracy is verified. Regardless of the specific representation of diagonal connections, the remaining steps and procedures of MAPES are identical to those described in this paper.
	
	\subsection{Virtual Ports and Impedance Matrix}
\begin{figure}[t]
		\begin{centering}
			\textsf{\includegraphics[width=1\columnwidth]{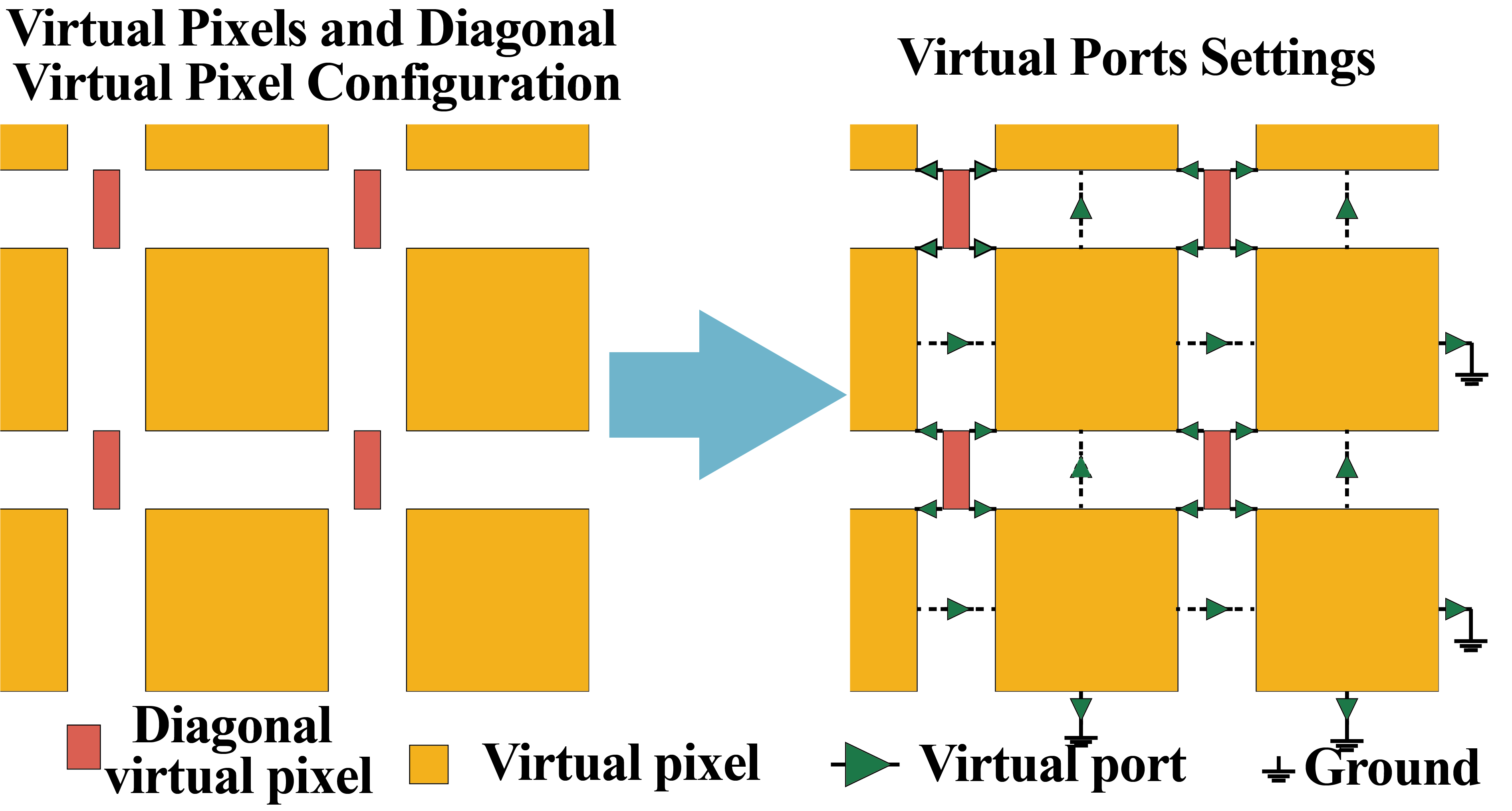}}
			\par\end{centering}
		\caption{Magnified view of the virtual pixel structure (left) and the insertion of virtual ports (right), including horizontal, vertical and diagonal virtual ports in the MAPES methodology.}
		\label{step3}
	\end{figure}

	With the virtual pixels and diagonal virtual pixels modeled over the design space, virtual ports are inserted as shown in  Fig. \ref{working_flow}(e)  and in the detailed view of Fig. \ref{step3}.
	
	The insertion of virtual ports is organized into three sets. The first set comprises virtual ports placed between adjacent virtual pixels to capture physical properties in the vertical and horizontal directions of original pixels. The second set contains virtual ports between diagonal virtual pixels and adjacent virtual pixels to characterize diagonal interactions of original pixels. The final set consists of virtual ports between virtual pixels and the ground. These ports facilitate the definition of I/O ports by capturing the coupling between pixels and ground. Typically, these ground ports are placed between all outer virtual pixels and the ground to allow flexible I/O placement.
	
	For multi-layer configurations, an optional additional set of virtual ports can be introduced. Extra virtual ports can be inserted between corresponding virtual pixels on adjacent layers to capture the electromagnetic properties of potential vias between layers, thereby extending the design space to include inter-layer connections that were not considered in previous works \cite{Karahan2023princeton,Chu2025ETH,Chu2025eth1,Guo2025princeton,Karahan2024princeton}.
	
	\begin{figure}[t]
		\begin{centering}
			\textsf{\includegraphics[width=0.8\columnwidth]{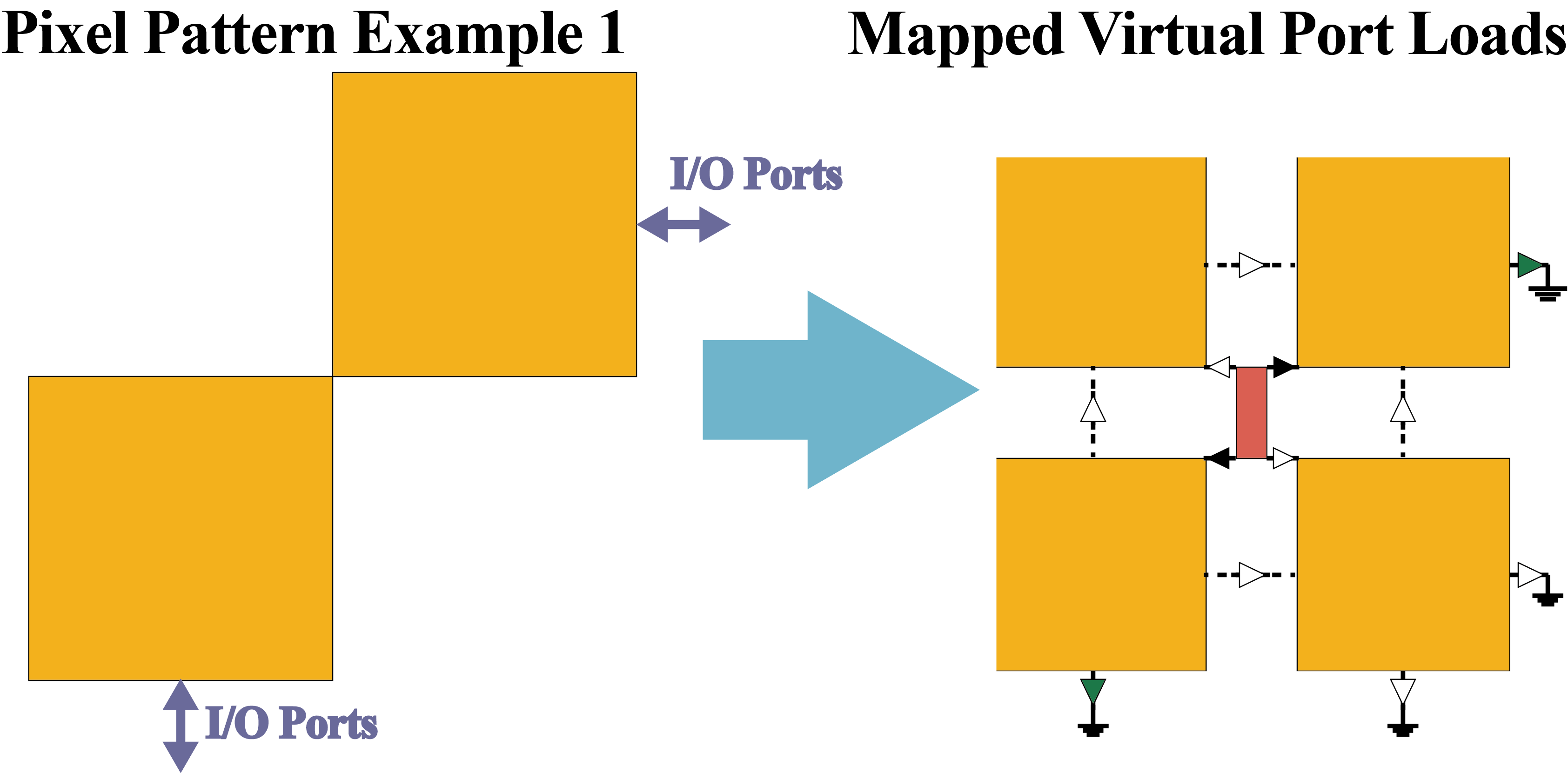}}
			\par\end{centering}
		
		\begin{centering}
			\textsf{\includegraphics[width=0.8\columnwidth]{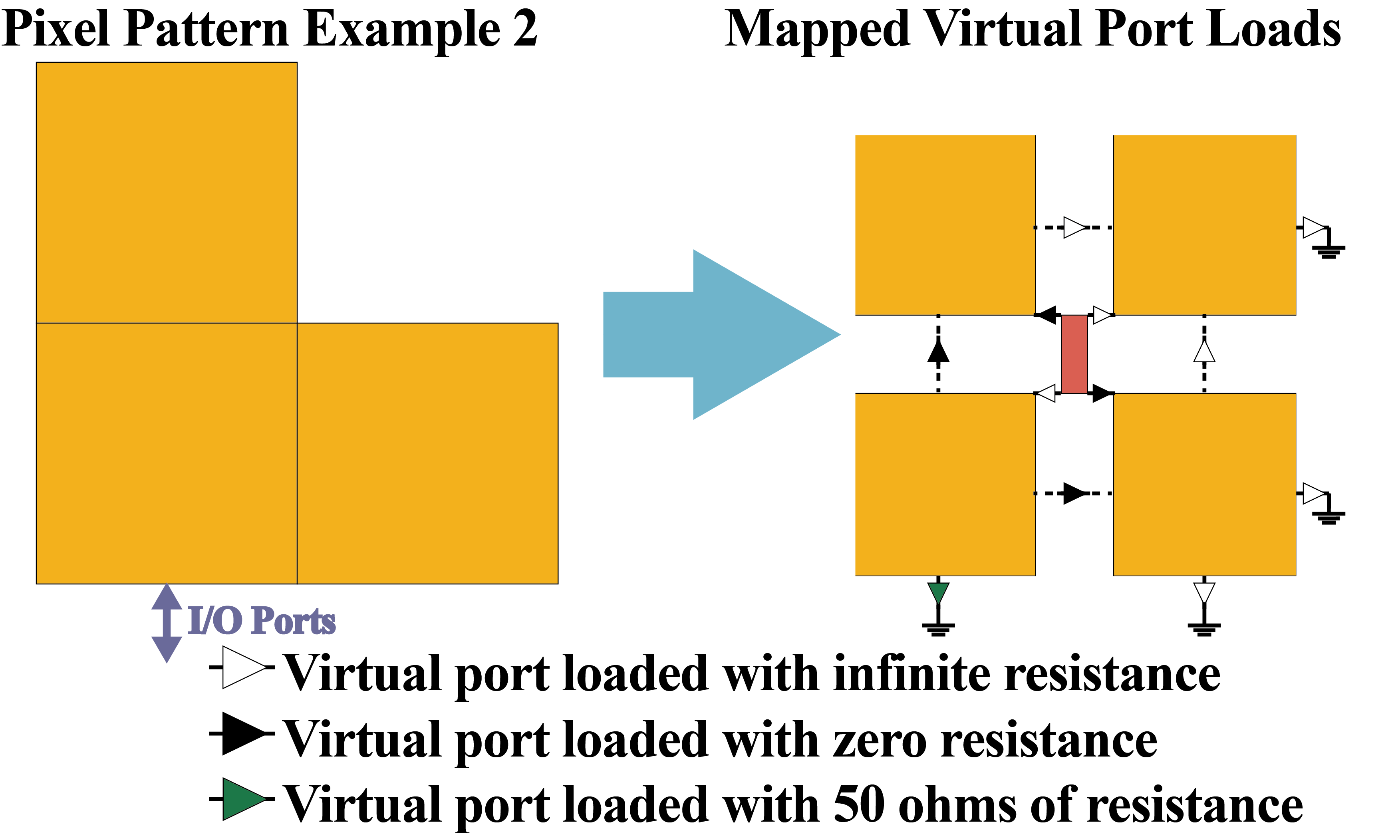}}
			\par\end{centering}
		
		\caption{Two representative examples of how the design space pixel patterns are mapped to the equivalent model in MAPES methodology.}
		\label{step5}
	\end{figure}

	For a design space with $L$ layers of $M \times N$ (row $\times$ column) pixels, without considering virtual ports between different layers, the total number of virtual ports, $Q$, is given by: 
	
	\begin{equation}
		Q=L(6MN-3M-3N+4),  \label{eq:1}
	\end{equation}
	while including the virtual ports between different layers for vias, the total number of virtual ports $Q$ is:
	\begin{equation}
		Q=L(6MN-3M-3N+4)+(L-1)MN.  \label{eq:2}
	\end{equation}
	
	With the inserted virtual ports, the RF properties across the entire design space can be comprehensively captured, as shown in Fig. \ref{working_flow} (f). Full-wave simulations can then be performed to obtain the impedance matrix, $\mathbf{Z}_{\mathrm{ALL}} \in \mathbb{C}^{Q\times Q}$, which includes all necessary information for calculating the performance of arbitrary input pixel patterns. It is worth noting that other equivalent matrices, such as the S-parameter matrix or admittance matrix, can also be used in MAPES if needed.
	
	\textcolor{black}{In practice, $\mathbf{Z}_{\mathrm{ALL}}$ is obtained from a single standard
		multiport discrete-port simulation: one discrete port is assigned at each of the $Q$
		virtual-port locations with a common 50\,$\Omega$ reference, and the resulting $Q\times Q$
		scattering matrix is converted to $\mathbf{Z}_{\mathrm{ALL}}$. The detailed procedure is given
		in Appendix~A.}
	
	This impedance matrix, $\mathbf{Z}_{\mathrm{ALL}}$, constitutes the entire prior dataset required by MAPES. Obtaining it requires $Q$ excitations of full-wave simulations using CST, HFSS, or other simulation software. The computational load is significantly reduced compared to the training data generation required by prior AI-assisted techniques, with full-wave simulation time or computational load being only about $1\%$ of that required by AI-assisted EM simulators.
	
	\subsection{Mapping Algorithm From Pixel Pattern to \(\mathbf{Z}_{L}\)}
	
	\begin{algorithm}
		
		\algnewcommand{\Input}{\item[\textbf{Input:}]}
		\algnewcommand{\Output}{\item[\textbf{Output:}]}
		
		\algrenewcommand{\algorithmicfor}{\textbf{for}}
		\algrenewcommand{\algorithmicdo}{}
		\algrenewcommand{\algorithmicend}{\textbf{end}}
		\algrenewcommand{\algorithmicif}{\textbf{if}}
		\algrenewcommand{\algorithmicthen}{\textbf{then}}
		\algrenewcommand{\algorithmicelse}{\textbf{else}}
		\caption{Pixel Pattern to Load Mapping Algorithm} \label{Mapping Algorithm}
		\begin{algorithmic}[1]
			\Input Pixel pattern $\mathbf{P}$ ($M \times N \times (2L-1)$ binary matrix)
			\Output Load matrix $\mathbf{Z}_{L}$ ($(Q-K) \times (Q-K)$ diagonal matrix)
			
			\State Initialize $\mathbf{Z}_{L}$ as zero matrix
			
			\Statex
			\State \textbf{Process 1: Process I/O ports}
			\For{all virtual ports between virtual pixels and ground}
			\If{the port is not an I/O port}
			\State Set corresponding load to infinite resistance (open circuit)
			\Else
			\State Skip mapping (reserved for I/O function)
			\EndIf
			\EndFor
			
			\Statex
			\State \textbf{Process 2: Process pixel patterns}
			\State Initialize all relevant virtual ports loads to zero resistance (short circuit)
			\For{all pixels $p_{i,j,l} \in \mathbf{P}$ (first $L$ layers)}
			\If{pixel $p_{i,j,l}$ is in absent state}
			\State Find all virtual ports connected to this absent pixel
			\State Reset these virtual ports loads from zero to infinite resistance (open circuit)
			\EndIf
			\EndFor
			
			\Statex
			\State \textbf{Process 3: Process vias between different layers}
			\For{all vias $p_{i,j,l} \in \mathbf{P}$ (last $L-1$ layers)}
			\If{via $p_{i,j,l}$ is in present state}
			\State Set corresponding virtual ports load to zero resistance (short circuit)
			\Else
			\State Set corresponding virtual ports load to infinite resistance (open circuit)
			\EndIf
			\EndFor
			
			\Statex
			\State \Return{Load matrix $\mathbf{Z}_{L}$}
		\end{algorithmic}
	\end{algorithm}
	With the prior dataset \(\mathbf{Z}_{\mathrm{ALL}}\) obtained from the preceding steps, we possess the information needed to compute the performance of arbitrary pixel patterns. The calculation is based on the microwave multiport system formed by all virtual ports.
	
	\begin{figure*}[t]
		\begin{centering}
			\textsf{\includegraphics[width=1.5\columnwidth]{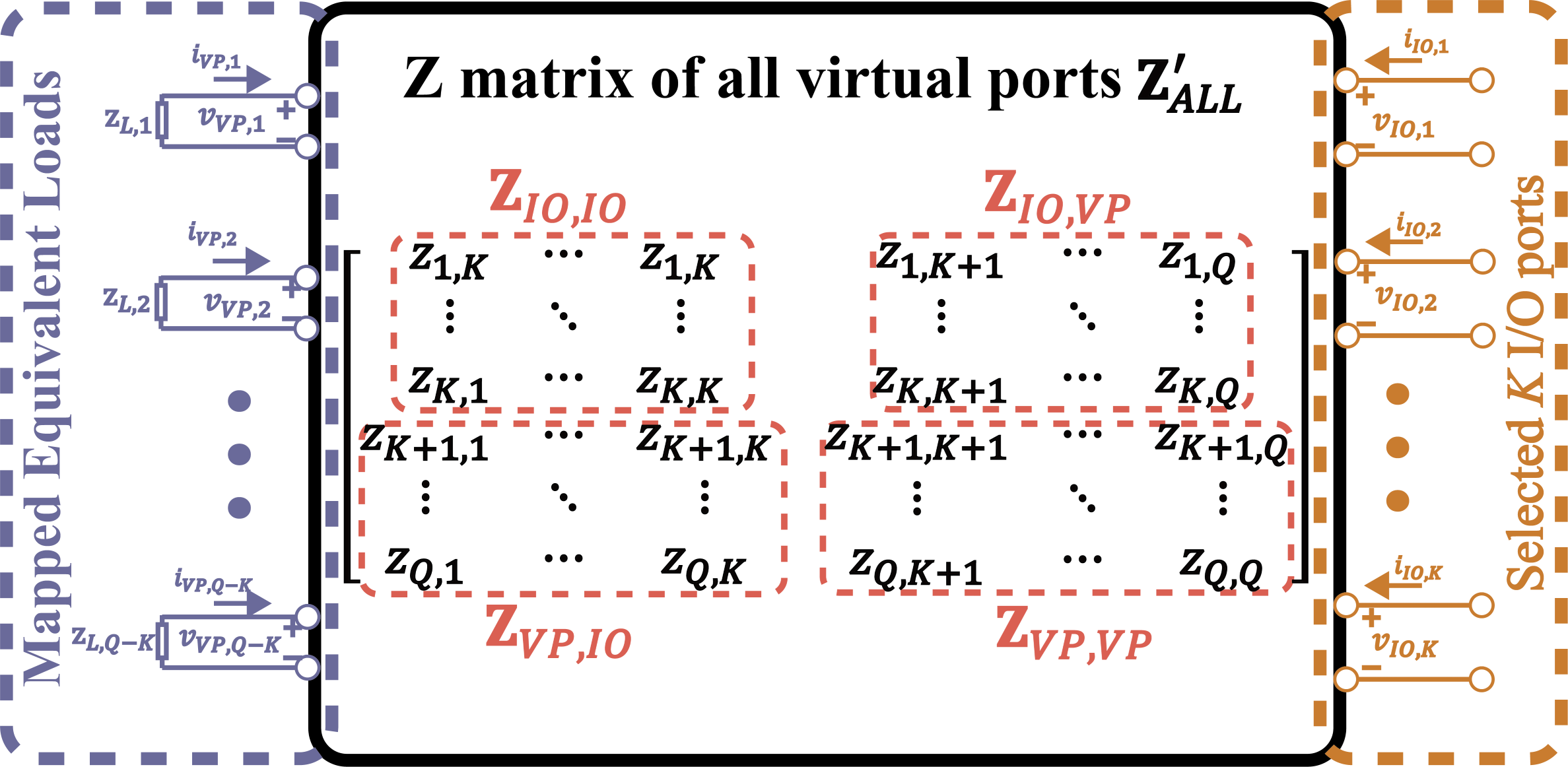}}
			\par\end{centering}
		\caption{Model of the multiport network for analytical calculation in the MAPES methodology. The connections between virtual pixels are modelled by the loads specified by $Z_L$ on the left side. The MAPES calculation process uses $Z_L$ with the Z matrix (middle) to determine the Z- or S-parameters of the I/O ports (right).}
		\label{multiport_theory}
	\end{figure*}

	For a design space with \(L\) layers of \(M\times N\) (rows \(\times\) columns) pixels and with possible vias between layers, any pixel configuration can be represented by an \(M\times N\times(2L-1)\) binary matrix \(\mathbf{P}\). The first \(L\) \(M\times N\) slices of \(\mathbf{P}\) represent presence/absence states of pixels on the \(L\) layers, while the following \((L-1)\) \(M\times N\) slices represent presence/absence states of vias between adjacent layers. Specifically, each entry \(p_{i,j,l}\), where \(1\le i\le M\), \(1\le j\le N\), and \(1\le l\le 2L-1\), takes the value 1 or 0: if \(1\le l\le L\), \(p_{i,j,l}=1\) indicates a pixel present at position \((i,j)\) on layer \(l\) (and \(0\) indicates absence); if \(L+1\le l\le 2L-1\), \(p_{i,j,l}=1\) indicates a via present at position \((i,j)\) between layer \((l-L)\) and \((l-L+1)\) (and \(0\) indicates absence). Note that a via can exist only between overlapping pixels on adjacent metal layers.
	
	Different connections among pixels can be represented as specific loads (short or open conditions) connected to the multiport system of virtual ports. To facilitate the derivation of the analytical formula in the next subsection, we propose a mapping algorithm that converts any pixel pattern into the corresponding load matrix \(\mathbf{Z}_{L}\). Assume \(K\) I/O ports are chosen among the \(Q\) virtual ports; then there are \(Q-K\) virtual ports to be assigned loads. Thus \(\mathbf{Z}_{L}=\mathrm{diag}(z_{L,1},\dots,z_{L,Q-K})\) is a \((Q-K)\times(Q-K)\) diagonal matrix whose diagonal entries represent the loads on the non-I/O virtual ports. An I/O port may be selected from any virtual port. However, for practical convenience we typically place I/O ports on outer pixels, and in this paper we restrict consideration to I/O ports located on outer pixels.
	
	The mapping in Algorithm~\ref{Mapping Algorithm} converts a given pixel matrix \(\mathbf{P}\) into a diagonal load matrix \(\mathbf{Z}_L(\mathbf{P})\) for all non-I/O virtual ports. Non-selected ground ports are terminated as open circuits, while the selected ground ports are reserved as external I/O terminals. For intra-layer and diagonal ports, loads are initialized as short circuits and then all ports connected with any absent pixel are reset to open circuits, thereby disconnecting the corresponding virtual pixel from its neighbors. For inter-layer via ports, each candidate via is modeled as a short circuit when present and an open circuit when absent (a finite via impedance can also be used when needed).
	
	After executing these three processes, the resulting diagonal matrix \(\mathbf{Z}_{L}\) gives the loads corresponding to the specific pixel pattern \(\mathbf{P}\). The proposed mapping algorithm is summarized in Algorithm \ref{Mapping Algorithm} for clarity. Two representative examples of the mapping algorithm are shown in Fig. \ref{step5}.

	\begin{table*}[t]
		\centering
		\caption{Summary of the four MAPES example configurations in Sec.~III.}
		\label{tab:ex_cfg}
		\setlength{\tabcolsep}{3.6pt}
		\renewcommand{\arraystretch}{1.15}
		\footnotesize
		
		\newcommand{\wCase}{0.12\textwidth}
		\newcommand{\ws}{0.28\textwidth}
		\newcommand{\wP}{0.07\textwidth}
		\newcommand{\wPix}{0.1\textwidth}
		\newcommand{\wArea}{0.1\textwidth}
		\newcommand{\wBand}{0.09\textwidth}
		\newcommand{\wQK}{0.13\textwidth}
		
		\begin{tabularx}{\textwidth}{
				C{\wCase} C{\ws} C{\wP} C{\wPix} C{\wArea} C{\wBand} C{\wQK}
			}
			\hline
			\textbf{Case (Sec.~III)} &
			\textbf{Platform / stackup} &
			\textbf{Design space}&
			\textbf{Pixel size} &
			\textbf{Design area} &
			\textbf{Band} &
			\textbf{\(Q\) / \(K\) / \textcolor{black}{\(T_{\mathbf{Z}_{\mathrm{ALL}}}\)}} \\
			\hline
			A: PCB (SL) & Rogers RO4003C (\(\varepsilon_r=3.55\), \(\tan\delta=0.0027\), \(h=0.203\)\,mm); single Cu layer; backside ground plane & \(16\times16\) & \(1.2\times1.2~\mathrm{mm}^2\) & \(19.2\times19.2~\mathrm{mm}^2\) & \(2\)--\(6\)\,GHz & 1444 / 2 / \textcolor{black}{10\,h\,1\,min} \\
			B: PCB (DL+via) & Two-layer PCB on Rogers RO4003C (each \(h=0.203\)\,mm); vias between the two pixel layers; bottom ground plane & \(16\times16\times3\) & \(0.61\times0.61~\mathrm{mm}^2\) & \(9.76\times9.76~\mathrm{mm}^2\) & \(2\)--\(6\)\,GHz& 3144 / 2 / \textcolor{black}{1\,d\,12\,h\,40\,min} \\
			C: CMOS 180\,nm (SL) & CMOS 180\,nm RF CMOS; top thick metal as design layer & \(17\times17\) & \(19\times19~\mu\mathrm{m}^2\) & \(323\times323~\mu\mathrm{m}^2\) & \(30\)--\(100\)\,GHz & 1636 / 8 / \textcolor{black}{17\,h\,16\,min} \\
			D: CMOS 65\,nm (DL+via) & CMOS 65\,nm RF process; two metal layers (top thick metal + Metal~8) with vias & \(13\times13\times3\) & \(23\times23~\mu\mathrm{m}^2\) & \(299\times299~\mu\mathrm{m}^2\) & \(30\)--\(100\)\,GHz & 2049 / 8 / \textcolor{black}{1\,d\,6\,h\,9\,min} \\
			\hline
		\end{tabularx}
		\vspace{6pt}
		\begin{minipage}{0.96\textwidth}
			\footnotesize
			SL: single layer; DL: double layer; \(Q\): virtual ports; \(K\): I/O ports.
			\textcolor{black}{\(T_{\mathbf{Z}_{\mathrm{ALL}}}\) denotes the one-time CST time-domain simulation time required to extract the prior matrix \(\mathbf{Z}_{\mathrm{ALL}}\). This extraction was performed on a standalone server with an AMD Ryzen 9 9950X3D CPU and an NVIDIA GeForce RTX 5090 GPU because it is more time- and memory-demanding than the subsequent MAPES evaluations. The server was not a high-performance computing cluster.}
		\end{minipage}
	\end{table*}
	
	\subsection{Analytical Formulas}
	With the impedance information of all virtual ports, \(\mathbf{Z}_{\mathrm{ALL}}\), obtained in Subsection II.B and the mapping algorithm that converts an arbitrary pixel pattern to the corresponding load matrix on the virtual ports in Subsection II.C, we have the necessary ingredients to derive analytical expressions for the performance of any pixel pattern at the chosen I/O ports based on microwave multiport theory \cite{Pozar2021} as shown in Fig. \ref{working_flow}(c).
	
	First, we restate the key definitions for better demonstration of the derivation below. Consider a design space with \(L\) layers of \(M\times N\) (rows \(\times\) columns) pixels, with possible vias between adjacent layers. Any pixel configuration can be represented by an \(M\times N\times(2L-1)\) binary matrix \(\mathbf{P}\): the first \(L\) \(M\times N\) slices indicate presence/absence of pixels on each layer, and the following \((L-1)\) slices indicate presence/absence of vias between adjacent layers. Select \(K\) I/O ports from the \(Q\) virtual ports; the remaining \(Q-K\) virtual ports are connected to loads and form the non-I/O virtual ports set. The corresponding load matrix of this set is the diagonal matrix \(\mathbf{Z}_{L}=\mathrm{diag}(z_{L,1},\dots,z_{L,Q-K})\), which depends on \(\mathbf{P}\).

	For convenience, we reorder the ports of \(\mathbf{Z}_{\mathrm{ALL}}\) so that the selected \(K\) I/O ports are located in the first $K \times K$ sub-matrix in  \(\mathbf{Z}_{\mathrm{ALL}}\); denote this reordered impedance matrix by \(\mathbf{Z}_{\mathrm{ALL}}'\), which can be partitioned as
	\begin{equation}
		\mathbf{Z}_{\mathrm{ALL}}'=\begin{bmatrix}
			\mathbf{Z}_{IO,IO} & \mathbf{Z}_{IO,VP} \\
			\mathbf{Z}_{VP,IO} & \mathbf{Z}_{VP,VP}
		\end{bmatrix},  \label{eq:3}
	\end{equation}
	where \(\mathbf{Z}_{IO,IO}\) is  the  \(K\times K\) impedance matrix among the selected $K$ virtual ports for I/O purpose, \(\mathbf{Z}_{VP,VP}\) is  the \((Q-K)\times(Q-K)\) impedance matrix among the unselected virtual ports, and \(\mathbf{Z}_{IO,VP}\) and \(\mathbf{Z}_{VP,IO}\) are the mutual-impedance submatrices between I/O ports and remaining virtual ports. Note that \(\mathbf{Z}_{\mathrm{ALL}}'\) is obtained from \(\mathbf{Z}_{\mathrm{ALL}}\) by a permutation of rows and columns and therefore requires no additional full-wave simulations.
	
	As illustrated in Fig. \ref{multiport_theory}, the virtual ports, the selected I/O ports, and the loads determined from the pixel pattern form a multiport network. Collect the currents at the I/O ports in the vector \(\mathbf{I}_{IO}=[i_{IO,1},\dots,i_{IO,K}]^{\mathrm{T}}\) and the voltages in \(\mathbf{V}_{IO}=[v_{IO,1},\dots,v_{IO,K}]^{\mathrm{T}}\). For the non-I/O (virtual) ports, collect the currents in \(\mathbf{I}_{VP}=[i_{VP,1},\dots,i_{VP,Q-K}]^{\mathrm{T}}\) and the voltages in \(\mathbf{V}_{VP}=[v_{VP,1},\dots,v_{VP,Q-K}]^{\mathrm{T}}\).
	
	By multiport network theory, the port voltages and currents are related by
	\begin{equation}
		\begin{bmatrix}
			\mathbf{V}_{IO} \\
			\mathbf{V}_{VP}
		\end{bmatrix}
		=
		\begin{bmatrix}
			\mathbf{Z}_{IO,IO} & \mathbf{Z}_{IO,VP} \\
			\mathbf{Z}_{VP,IO} & \mathbf{Z}_{VP,VP}
		\end{bmatrix}
		\begin{bmatrix}
			\mathbf{I}_{IO} \\
			\mathbf{I}_{VP}
		\end{bmatrix}.
		\label{eq:4}
	\end{equation}
	Besides, as in the left side of Fig. \ref{multiport_theory}, for the non-I/O ports connected to loads, the voltage-current relation is also given by
	\begin{equation}
		\mathbf{V}_{VP} = -\mathbf{Z}_{L}(\mathbf{P}) \mathbf{I}_{VP},
		\label{eq:5}
	\end{equation}
	where \(\mathbf{Z}_{L}(\mathbf{P})\) denotes the diagonal load-impedance matrix mapped from the pixel pattern \(\mathbf{P}\).
	
	Combining \eqref{eq:4} and \eqref{eq:5} and eliminating the virtual port currents yields
	\begin{equation}
		\mathbf{V}_{IO} =\left[\mathbf{Z}_{IO,IO}-\mathbf{Z}_{IO,VP} (\mathbf{Z}_{L}(\mathbf{P})+\mathbf{Z}_{VP,VP})^{-1}\mathbf{Z}_{VP,IO} \right]\mathbf{I}_{IO}.
		\label{eq:6}
	\end{equation}
	
	Equation \eqref{eq:6} provides the effective multiport relation at the selected I/O ports for the pixel pattern \(\mathbf{P}\). In other words, for a given pixel pattern \(\mathbf{P}\), the impedance matrix observed at the I/O ports, denoted \(\mathbf{Z}_{\mathrm{MAPES}}(\mathbf{P})\in\mathbb{C}^{K\times K}\), is obtained analytically as
	\begin{equation}
		\mathbf{Z}_{\mathrm{MAPES}}(\mathbf{P}) =\mathbf{Z}_{IO,IO}-\mathbf{Z}_{IO,VP} (\mathbf{Z}_{L}(\mathbf{P})+\mathbf{Z}_{VP,VP})^{-1}\mathbf{Z}_{VP,IO}.
		\label{eq:7}
	\end{equation}
	
	Therefore, for any pixel pattern \(\mathbf{P}\) we can efficiently compute the corresponding impedance matrix seen at the I/O ports using \eqref{eq:7}. From \(\mathbf{Z}_{\mathrm{MAPES}}(\mathbf{P})\) one can straightforwardly convert to equivalent representations such as scattering (S-) parameters or admittance matrices as required. The submatrices \(\mathbf{Z}_{IO,IO}\), \(\mathbf{Z}_{IO,VP}\), \(\mathbf{Z}_{VP,VP}\), and \(\mathbf{Z}_{VP,IO}\) are obtained from the prior one-time full-wave simulations in Fig. \ref{working_flow} and are valid for all pixel patterns \(\mathbf{P}\).
	
	\section{MAPES Examples}
	
\textcolor{black}{This section validates MAPES using four representative configurations: single-layer PCB, double-layer PCB with vias, CMOS 180\,nm single-layer, and CMOS 65\,nm double-layer with vias. Their key parameters are summarized in Table~\ref{tab:ex_cfg}. For each case, \(\mathbf{Z}_{\mathrm{ALL}}\) is extracted once and reused to evaluate pixel patterns analytically by~\eqref{eq:7}. Representative S-parameter results are compared with CST simulations and, for PCB cases, measurements.}

\textcolor{black}{For the PCB examples, the I/O ports are located at the left/right edges of the
	pixelated region and connected to SMA connectors through 50-$\Omega$ microstrip access
	lines. The transmission line widths are 0.4\,mm and 0.85\,mm for the single-layer and double-layer PCB cases, respectively. The final
	measurement reference planes have been de-embedded at the edges of the pixelated region. Thus, the CST, MAPES,
	and measured results use the same reference planes. For the CMOS examples, the I/O ports are
	directly defined at the selected outer-edge metal ports, without any coaxial/SMA interface.}

	\subsection{Single-Layer PCB Example}
	
	\begin{figure*}[t]
		\centering
		\begin{minipage}[b]{0.35\textwidth}
			\vspace{0pt}
			\centering
			\includegraphics[width=0.95\linewidth]{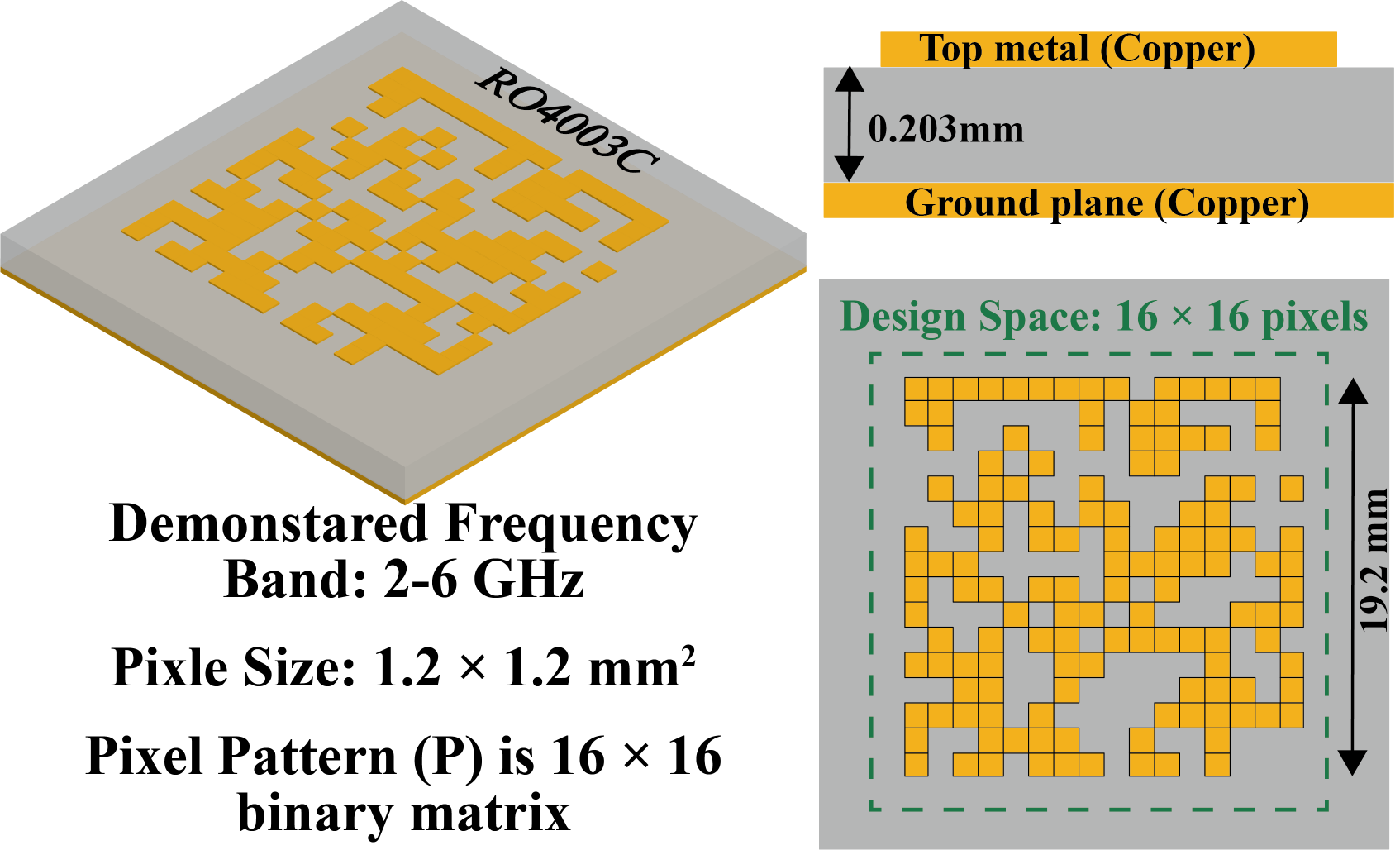}\\
			(a)
		\end{minipage}
		\hfill
		\begin{minipage}[b]{0.35\textwidth}
			\vspace{0pt}
			\centering
			\includegraphics[width=0.95\linewidth]{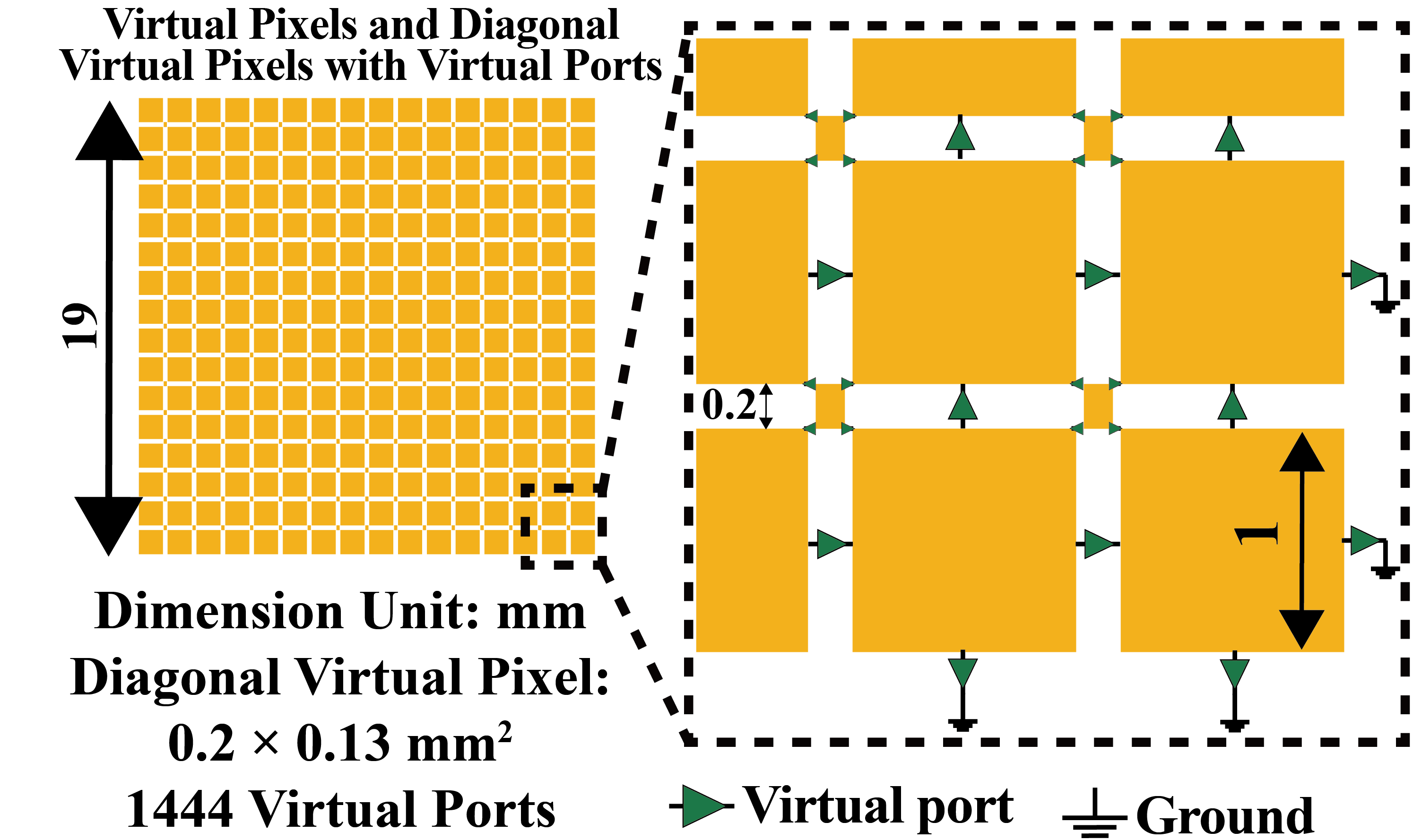}\\
			(b)
		\end{minipage}
		\hfill
		\begin{minipage}[b]{0.14\textwidth}
			\vspace{0pt}
			\centering
			\includegraphics[width=0.55\linewidth]{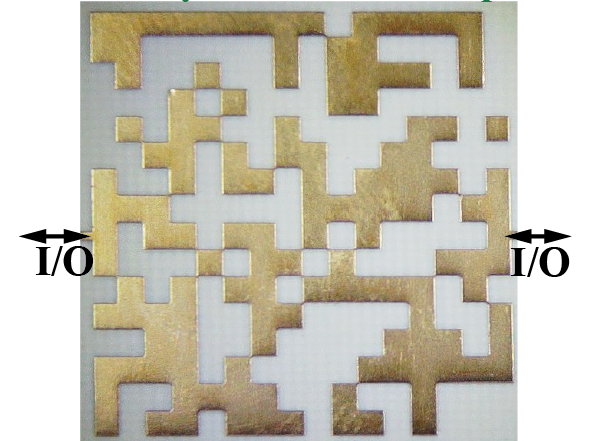}
			\includegraphics[width=0.95\linewidth]{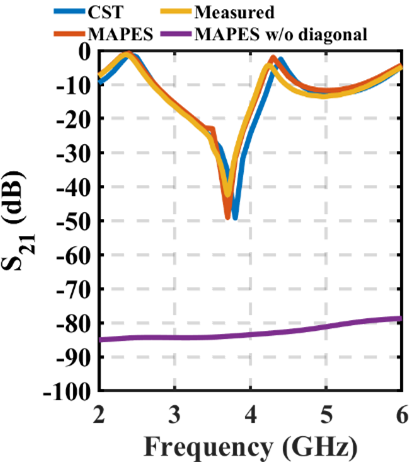}\\
			(c)
		\end{minipage}
		\hfill
		\begin{minipage}[b]{0.14\textwidth}
			\vspace{0pt}
			\centering
			\includegraphics[width=0.55\linewidth]{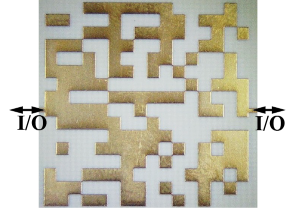}
			\includegraphics[width=0.95\linewidth]{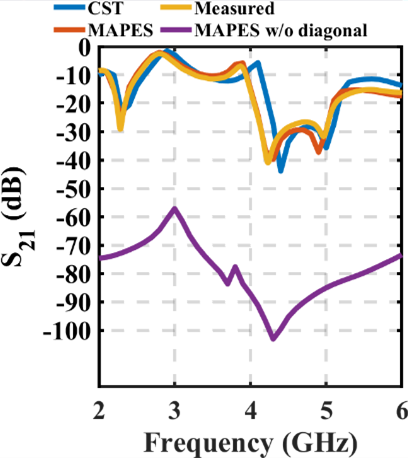}\\
			(d)
		\end{minipage}
		\caption{\textcolor{black}{PCB single-layer validation. (a) Pixel design space and physical stackup. (b) Virtual-pixel and virtual-port model. (c),(d) Two representative pixel patterns with prototype images and S-parameter comparison among CST, MAPES, MAPES without diagonal virtual pixels, and measured results.}}
		\label{fig:PCB1_all}
	\end{figure*}

	\begin{figure*}[t]
		\centering
		\begin{minipage}[b]{0.32\textwidth}
			\vspace{0pt}
			\centering
			\includegraphics[width=0.95\linewidth]{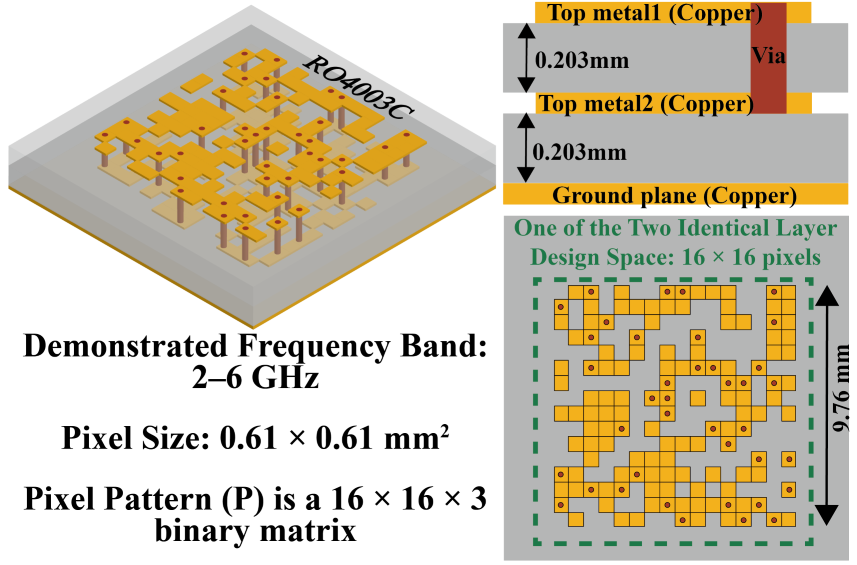}\\
			(a)
		\end{minipage}
		\hfill
		\begin{minipage}[b]{0.32\textwidth}
			\vspace{0pt}
			\centering
			\includegraphics[width=0.95\linewidth]{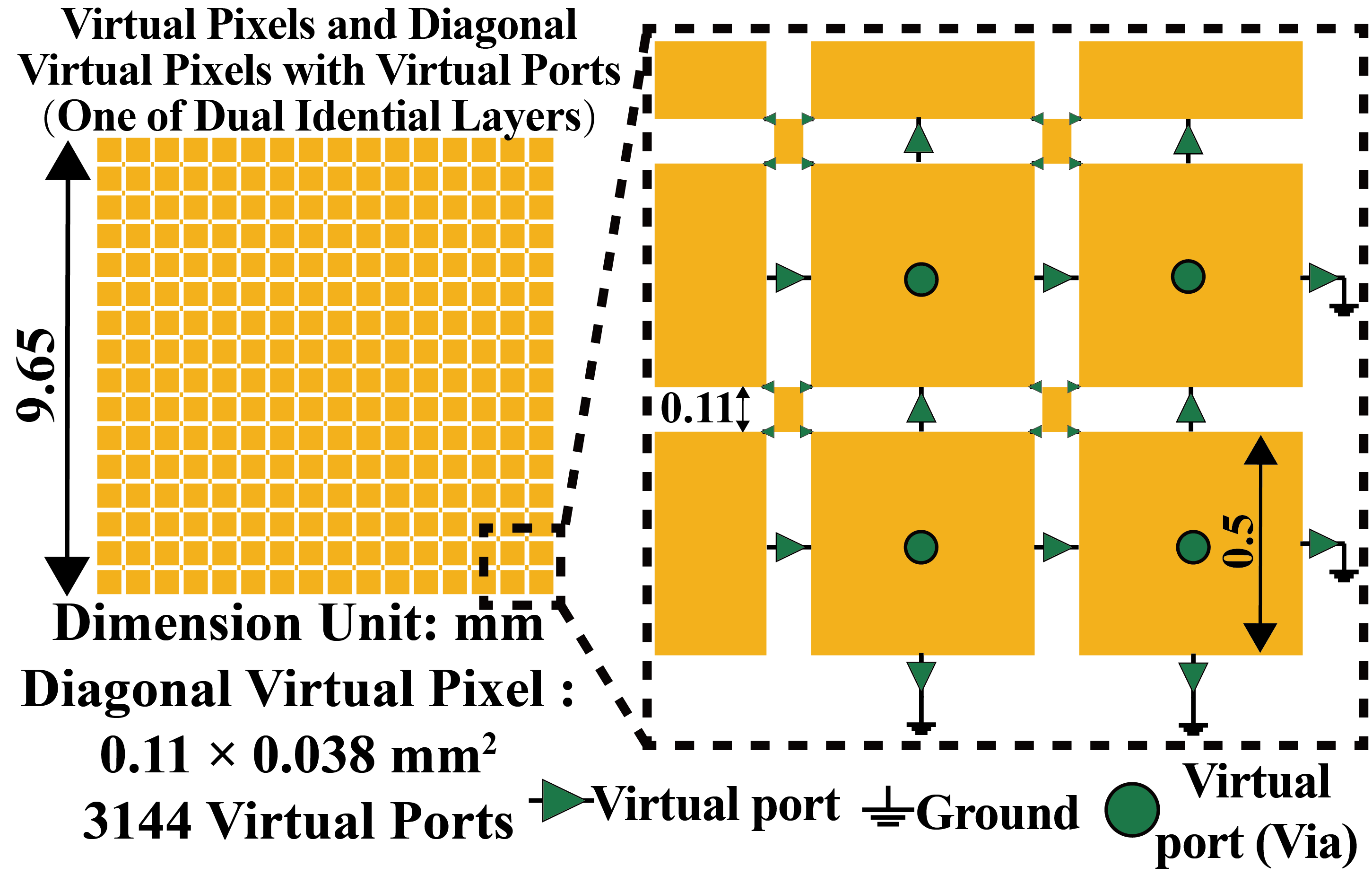}\\
			(b)
		\end{minipage}
		\hfill
		\begin{minipage}[b]{0.17\textwidth}
			\vspace{0pt}
			\centering
			\includegraphics[width=0.95\linewidth]{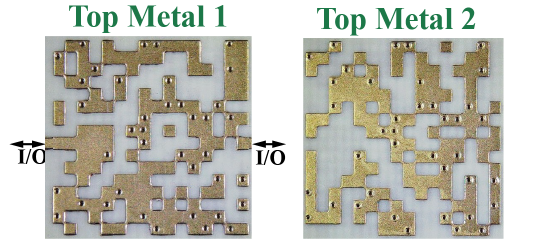}
			\vskip 0pt
			\includegraphics[width=0.99\linewidth]{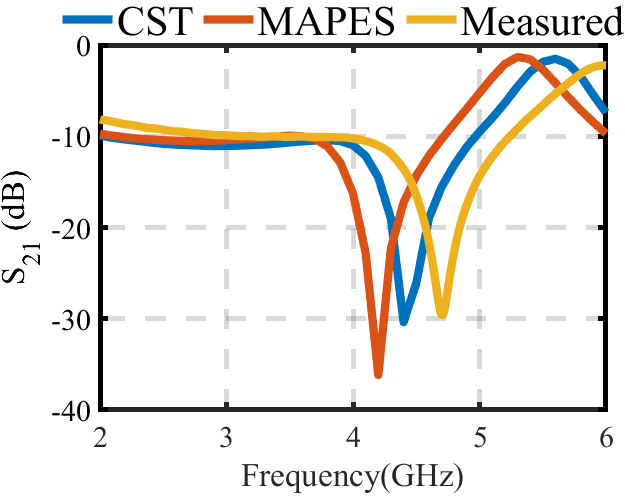}\\
			(c)
		\end{minipage}
		\hfill
		\begin{minipage}[b]{0.17\textwidth}
			\vspace{0pt}
			\centering
			\includegraphics[width=0.95\linewidth]{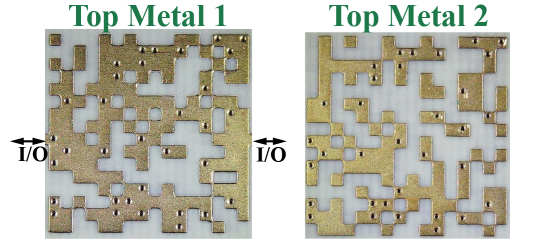}
			\vskip 0pt
			\includegraphics[width=0.99\linewidth]{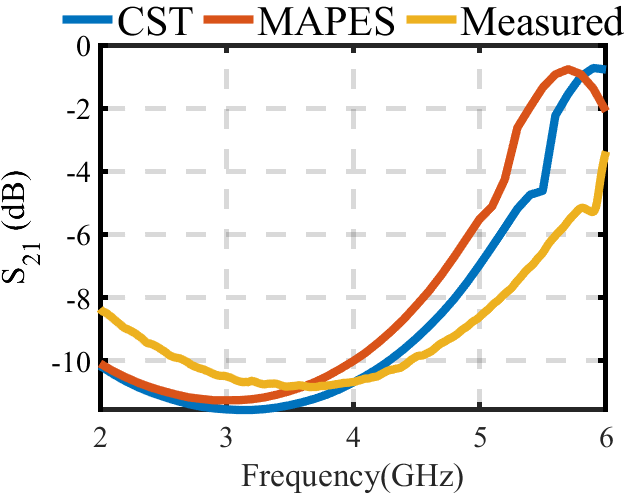}\\
			(d)
		\end{minipage}
		\caption{\textcolor{black}{PCB double-layer validation. (a) Pixel design space with via interconnection. (b) Virtual-pixel and virtual-port model for both layers. (c),(d) Two representative dual-layer pixel patterns with prototype images and S-parameter comparison between CST, MAPES, and measured results.}}
		\label{fig:PCB2_all}
	\end{figure*}

	\begin{figure*}[t]
		\centering
		\begin{minipage}[b]{0.34\textwidth}
			\vspace{0pt}
			\centering
			\includegraphics[width=0.95\linewidth]{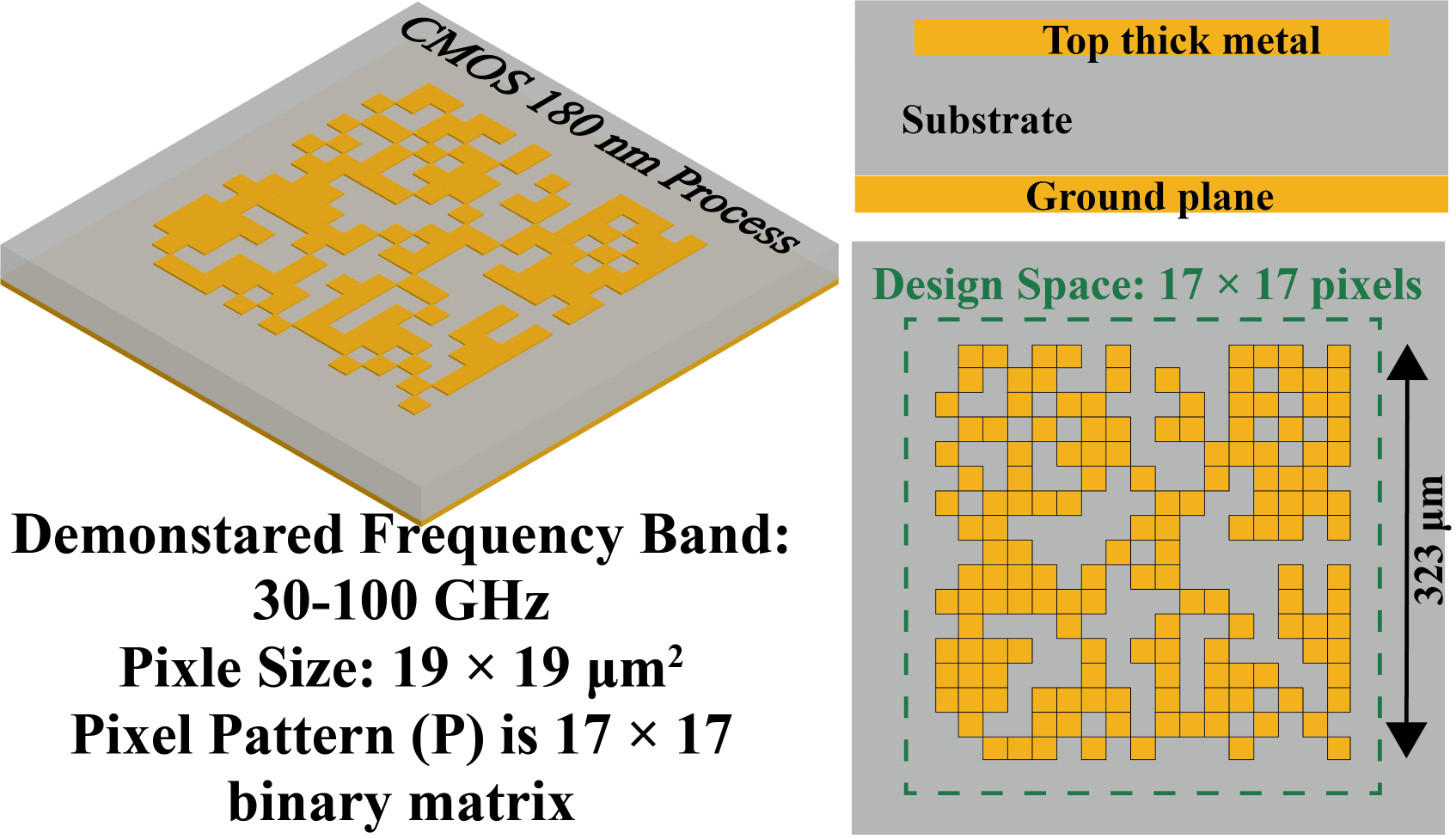}\\
			(a)
		\end{minipage}
		\hfill
		\begin{minipage}[b]{0.34\textwidth}
			\vspace{0pt}
			\centering
			\includegraphics[width=0.95\linewidth]{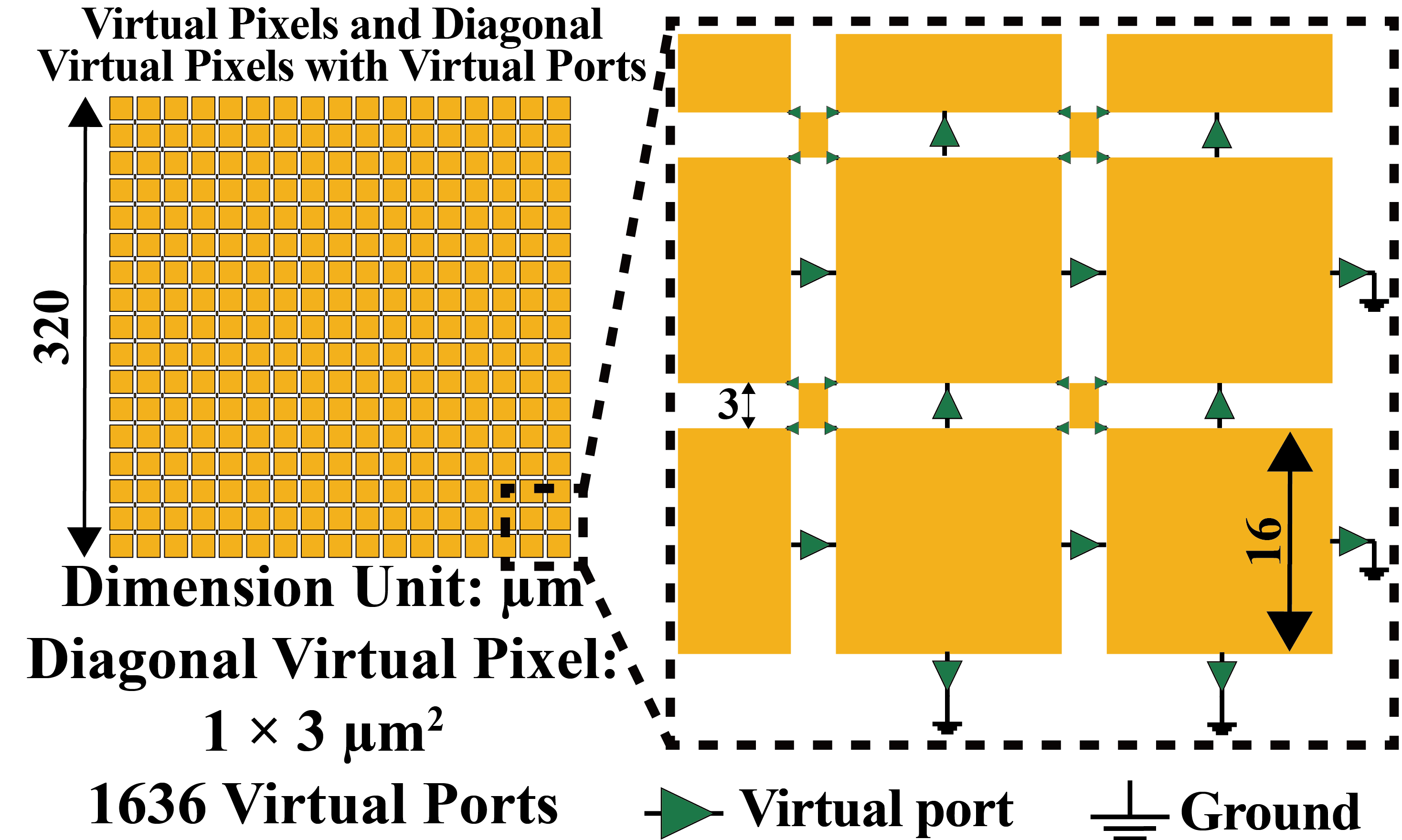}\\
			(b)
		\end{minipage}
		\hfill
		\begin{minipage}[b]{0.15\textwidth}
			\vspace{0pt}
			\centering
			\includegraphics[width=0.45\linewidth]{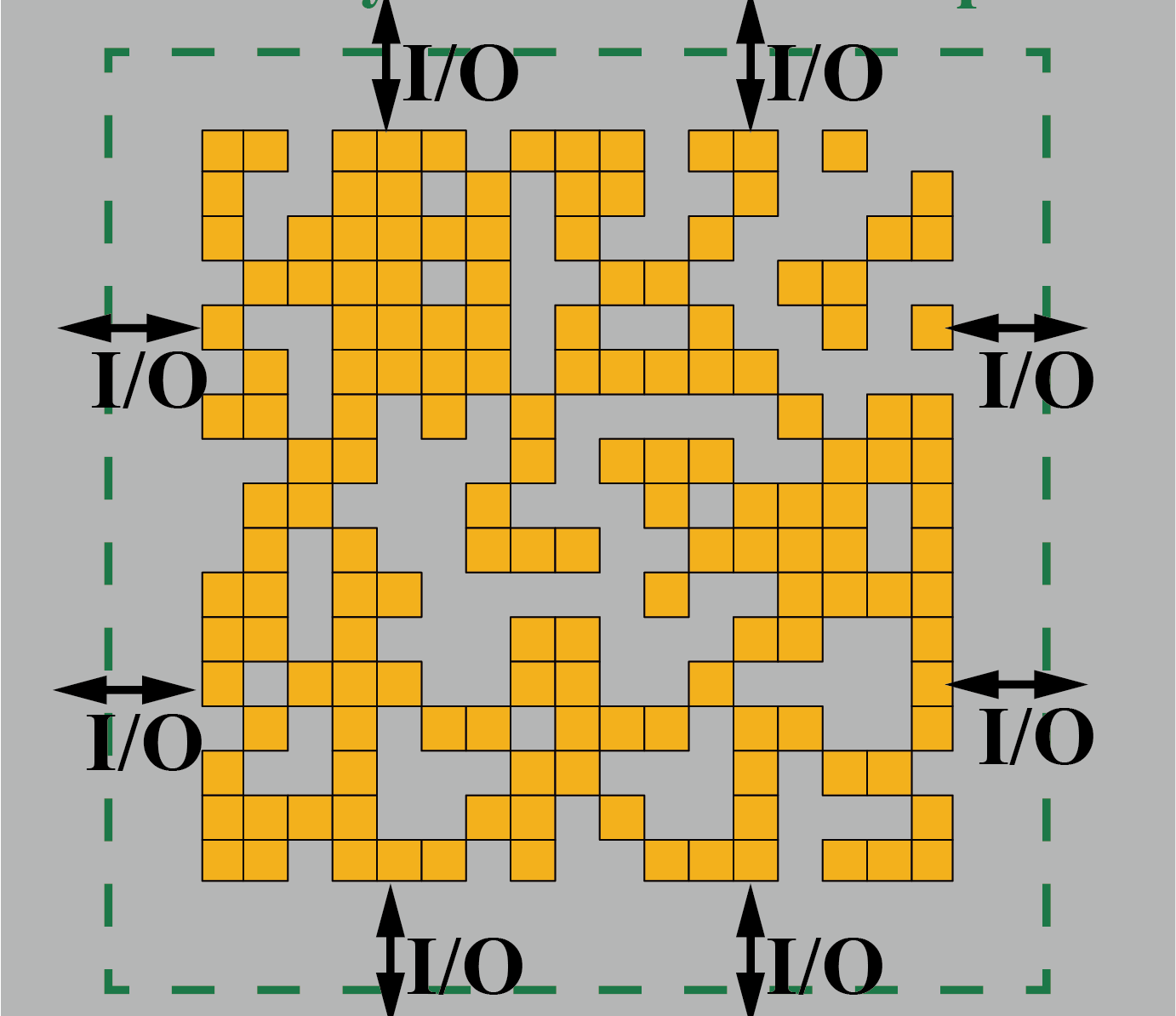}
			\vskip 0pt
			\includegraphics[width=0.99\linewidth]{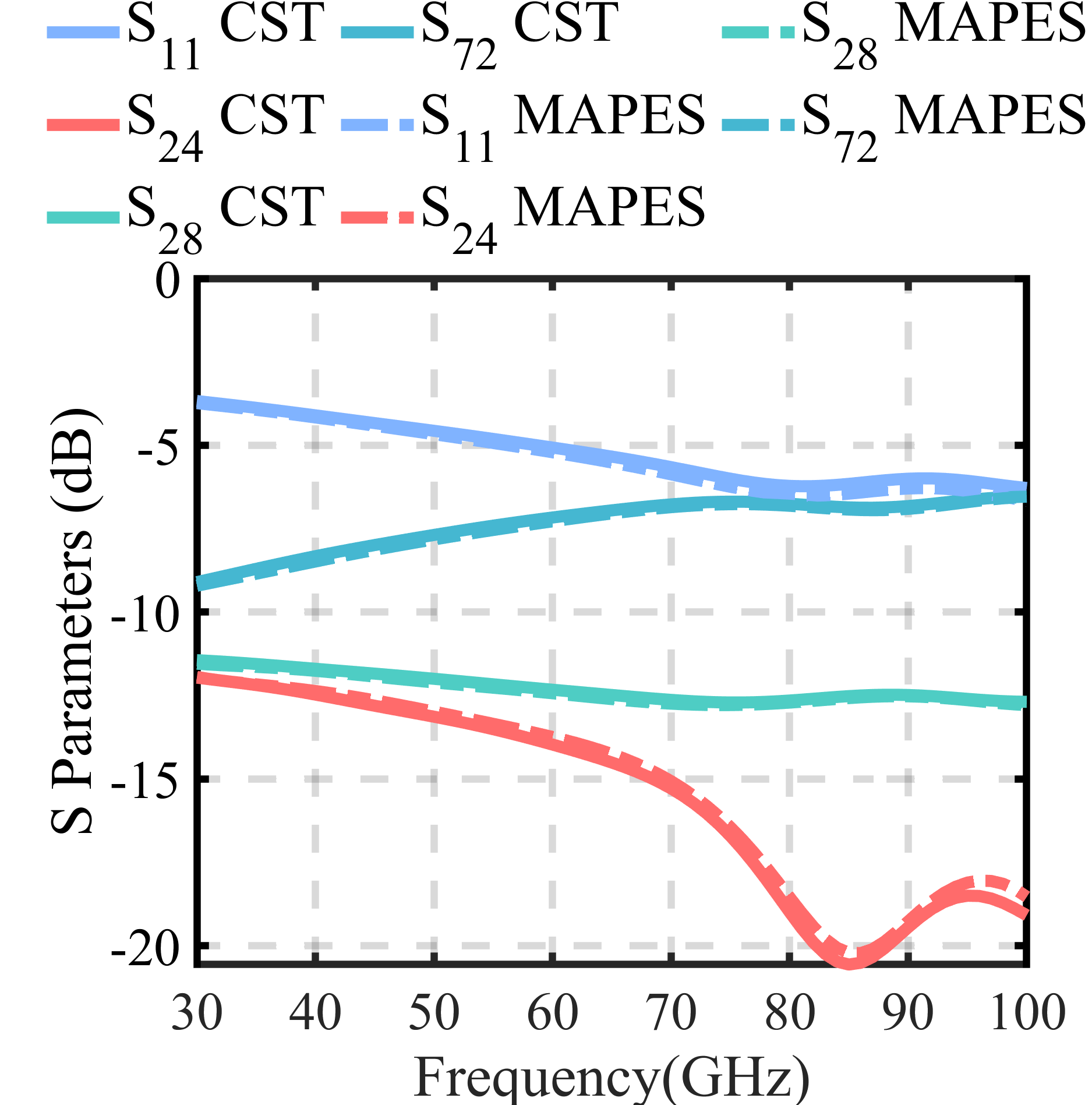}\\
			(c)
		\end{minipage}
		\hfill
		\begin{minipage}[b]{0.15\textwidth}
			\vspace{0pt}
			\centering
			\includegraphics[width=0.45\linewidth]{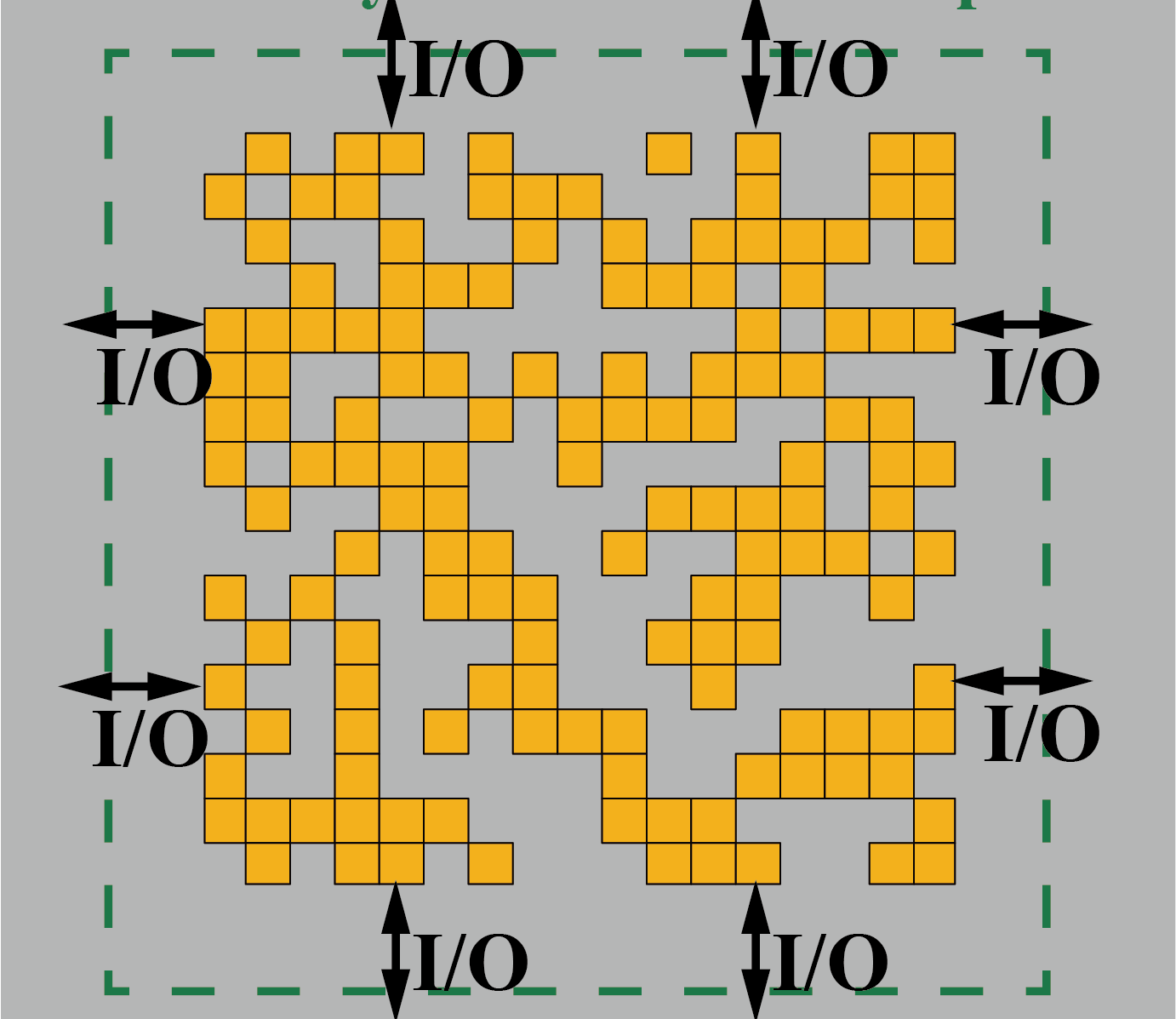}
			\vskip 0pt
			\includegraphics[width=0.99\linewidth]{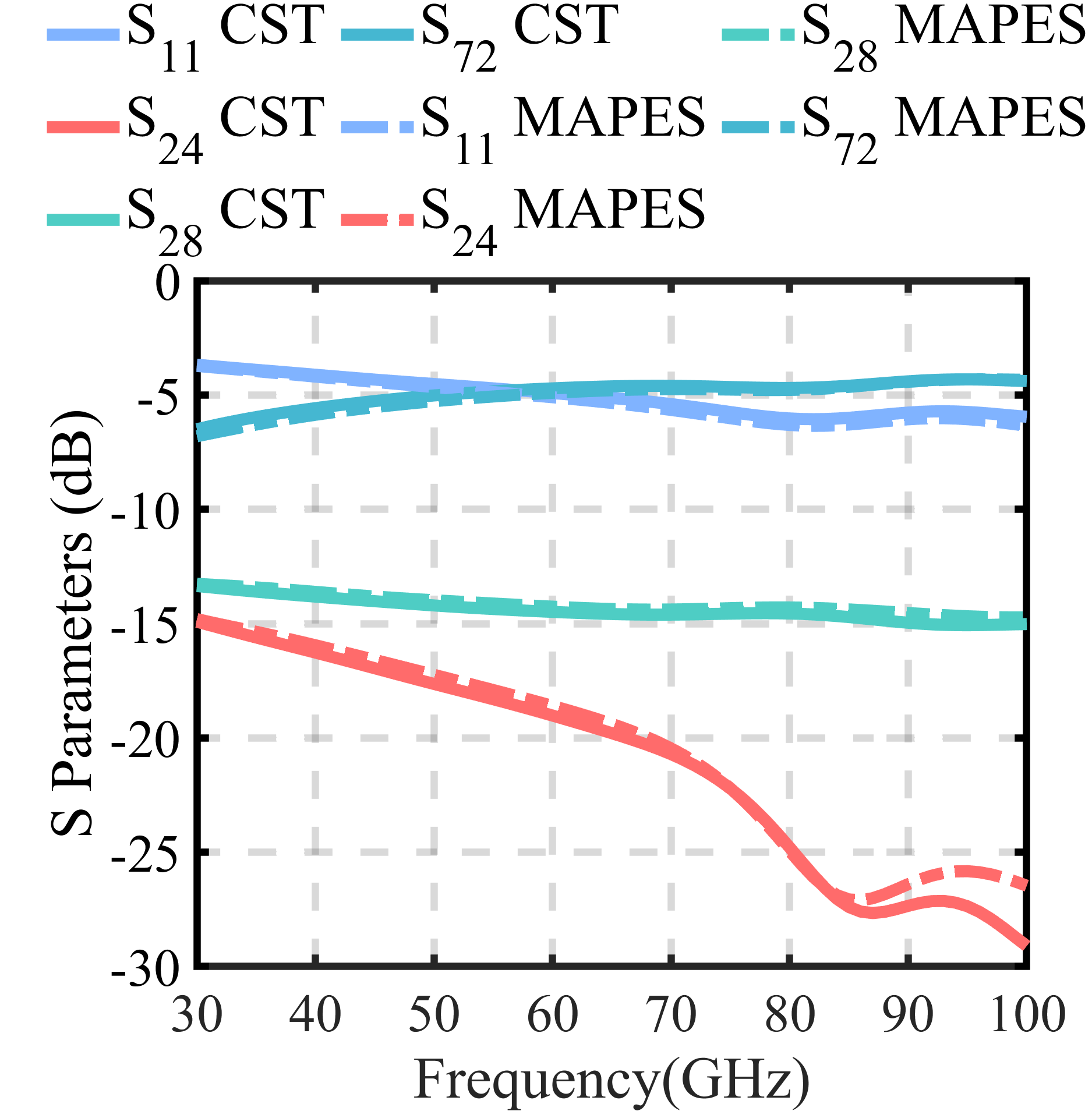}\\
			(d)
		\end{minipage}
		\caption{\textcolor{black}{CMOS 180\,nm single-layer validation. (a) Pixel design space on the top metal. (b) Virtual-pixel and virtual-port model. (c),(d) Two representative random pixel configurations with S-parameter comparison between CST and MAPES.}}
		\label{fig:CMOS180_all}
	\end{figure*}
	
	\begin{figure*}[t]
		\centering
		\begin{minipage}[b]{0.32\textwidth}
			\vspace{0pt}
			\centering
			\includegraphics[width=0.95\linewidth]{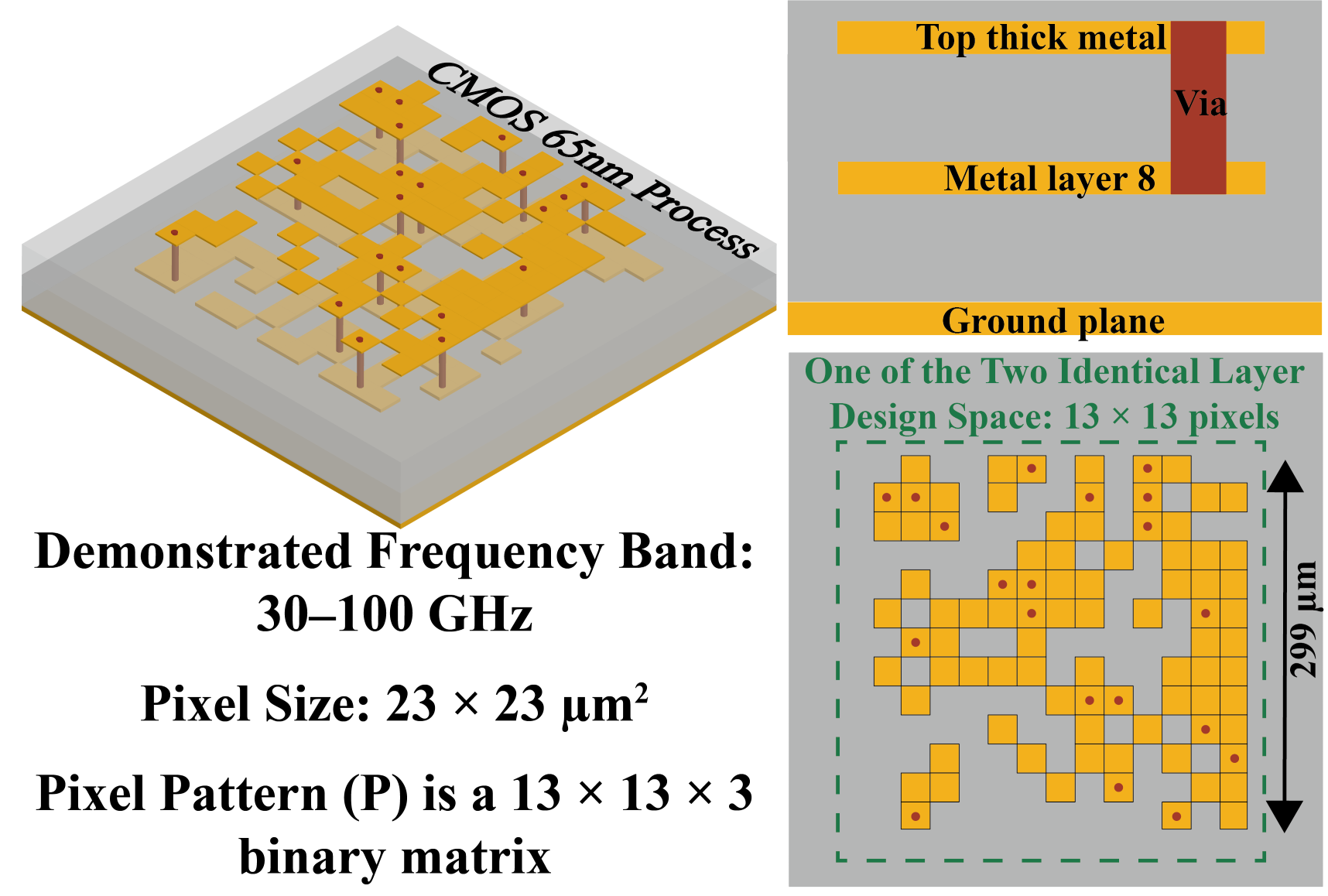}\\
			(a)
		\end{minipage}
		\hfill
		\begin{minipage}[b]{0.32\textwidth}
			\vspace{0pt}
			\centering
			\includegraphics[width=0.95\linewidth]{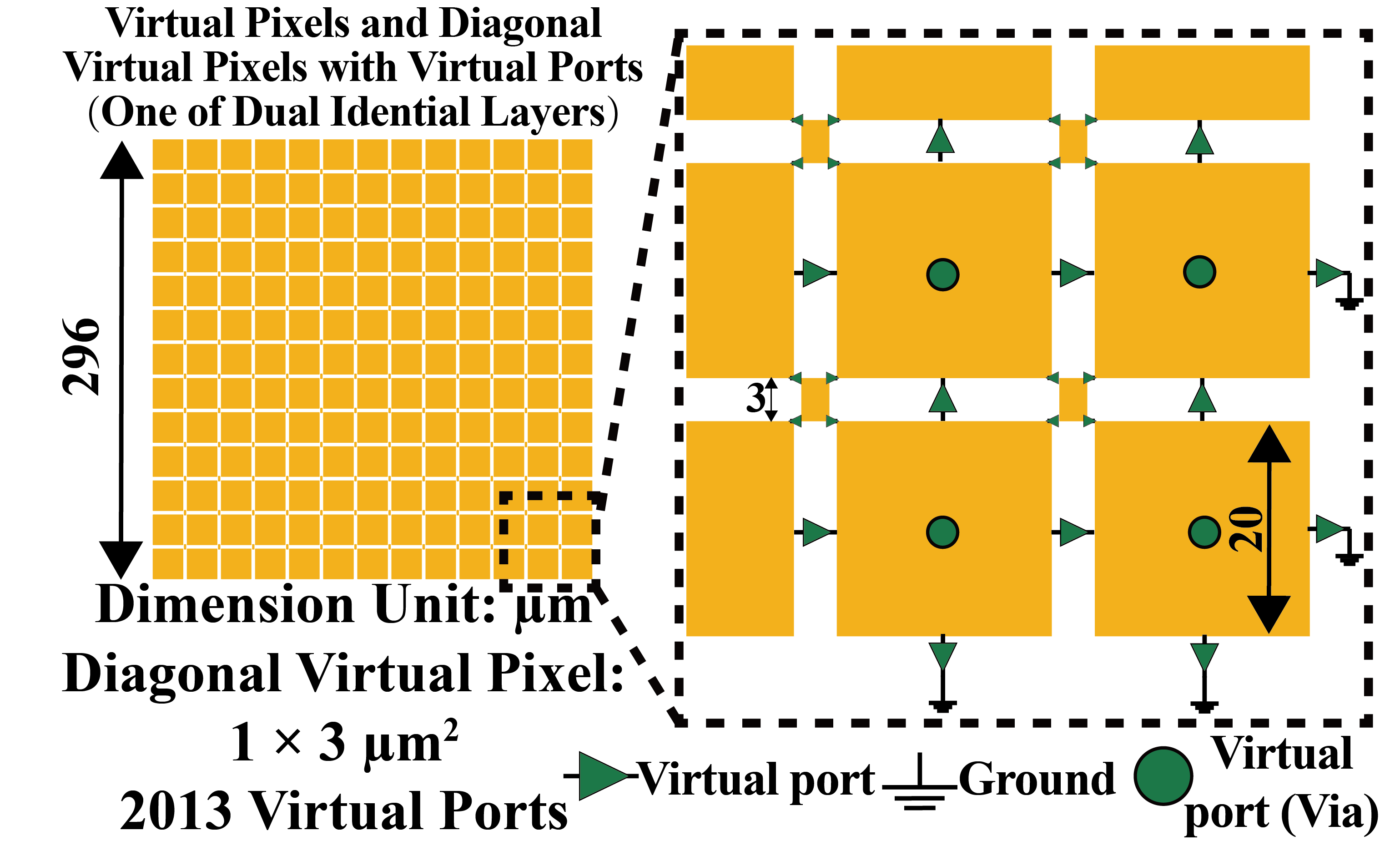}\\
			(b)
		\end{minipage}
		\hfill
		\begin{minipage}[b]{0.17\textwidth}
			\vspace{0pt}
			\centering
			\includegraphics[width=0.85\linewidth]{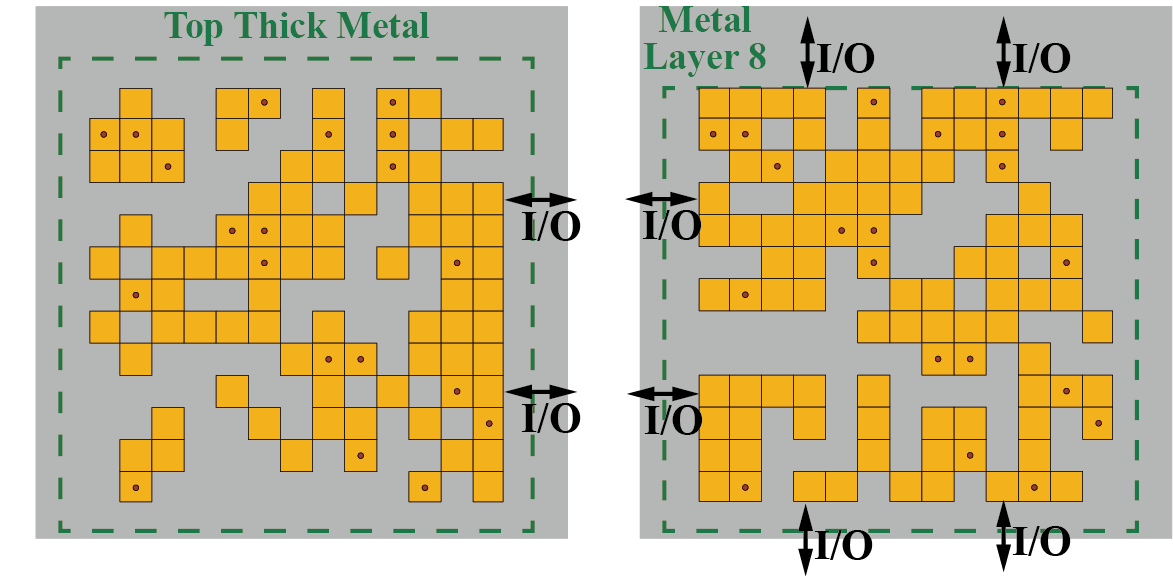}
			\vskip 0pt
			\includegraphics[width=0.95\linewidth]{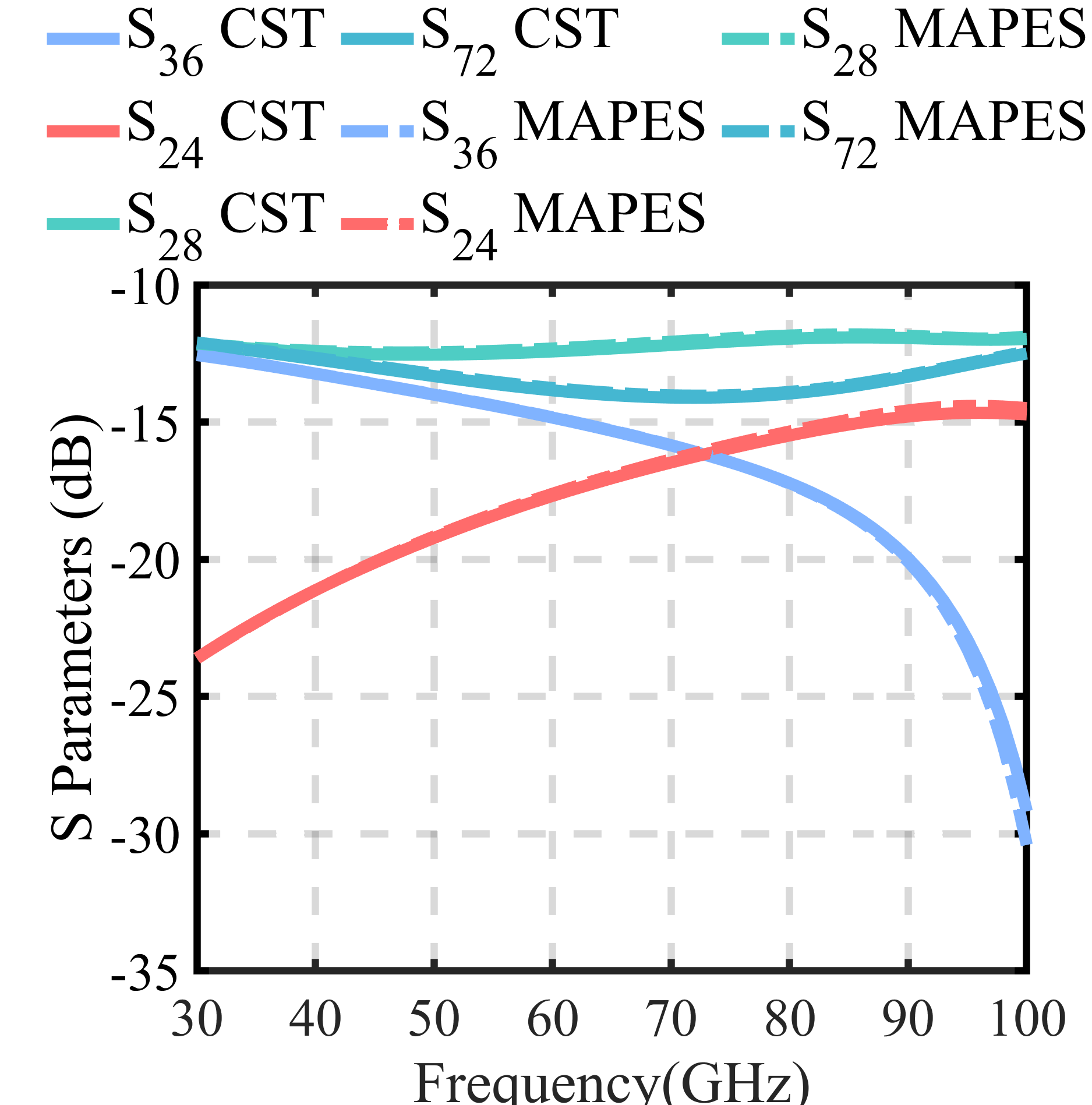}\\
			(c)
		\end{minipage}
		\hfill
		\begin{minipage}[b]{0.17\textwidth}
			\vspace{0pt}
			\centering
			\includegraphics[width=0.85\linewidth]{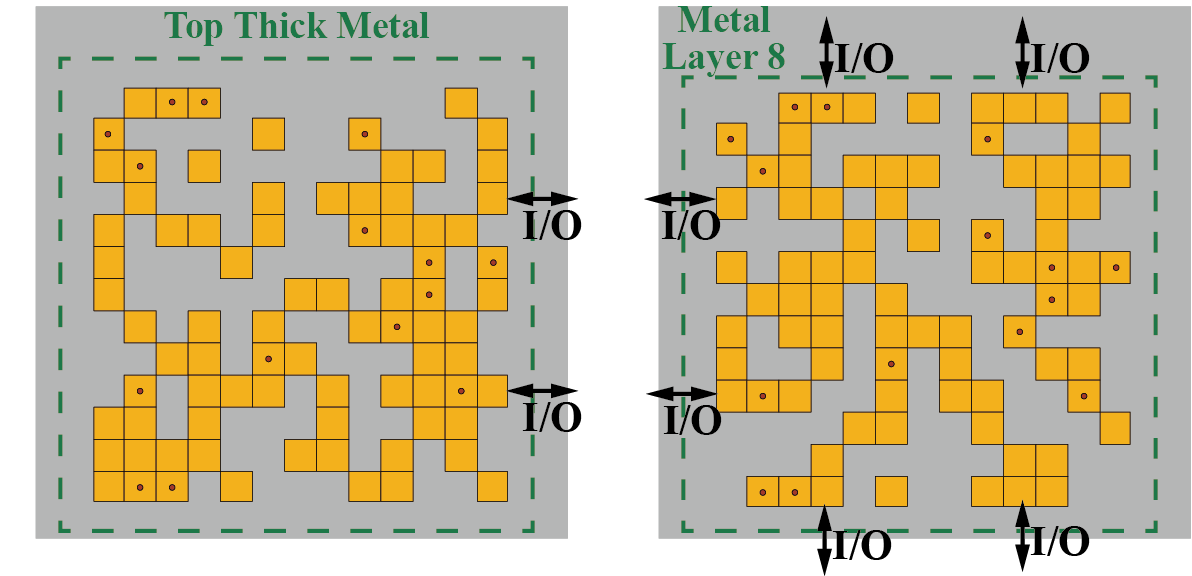}
			\vskip 0pt
			\includegraphics[width=0.95\linewidth]{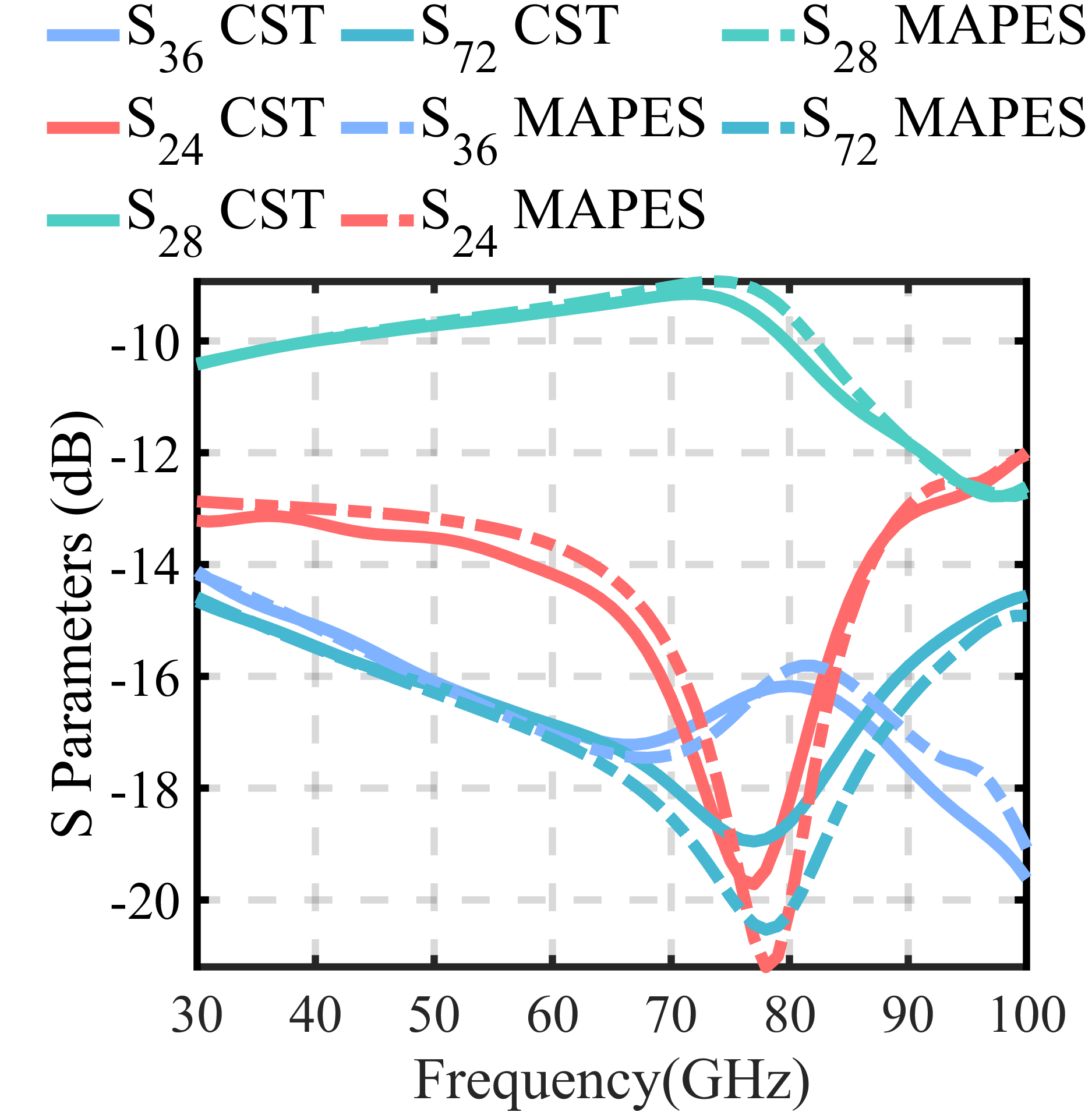}\\
			(d)
		\end{minipage}
		\caption{\textcolor{black}{CMOS 65\,nm double-layer validation. (a) Pixel design space with via interconnection. (b) Virtual-pixel and virtual-port model for both layers. (c),(d) Two representative dual-layer pixel configurations with S-parameter comparison between CST and MAPES.}}
		\label{fig:CMOS65_all}
	\end{figure*}

\textcolor{black}{The single-layer PCB setup and virtual-port construction are shown in Fig.~\ref{fig:PCB1_all}(a) and (b), with parameters summarized in Table~\ref{tab:ex_cfg}. After the one-time extraction of \(\mathbf{Z}_{\mathrm{ALL}}\), two representative random patterns with \(K=2\) outer-edge I/O ports are evaluated, as shown in Fig.~\ref{fig:PCB1_all}(c) and (d). MAPES agrees well with both CST and measurement over the full band.}

		\textcolor{black}{To validate diagonal virtual pixels, MAPES was evaluated with and without them for two random PCB single-layer patterns, also shown in  Fig.~\ref{fig:PCB1_all}(c) and (d). Removing the diagonal virtual pixels degrades the agreement with CST because diagonal coupling paths are no longer captured, confirming their importance in contiguous pixel layouts.}

	\subsection{Double-Layer PCB Example}

\textcolor{black}{The double-layer PCB configuration with vias is shown in Fig.~\ref{fig:PCB2_all}(a) and (b), with parameters summarized in Table~\ref{tab:ex_cfg}. After extracting \(\mathbf{Z}_{\mathrm{ALL}}\), two representative dual-layer patterns with \(K=2\) I/O ports are evaluated in Fig.~\ref{fig:PCB2_all}(c) and (d). The MAPES-computed \(S_{21}\) responses agree well with CST and measurements, demonstrating that MAPES captures both intra-layer and inter-layer coupling. The residual measurement mismatch is mainly attributed to assembly tolerances between the two PCB layers.}
	
	\subsection{CMOS 180 nm Single-Layer Example}
	
\textcolor{black}{The CMOS 180\,nm single-layer configuration is shown in Fig.~\ref{fig:CMOS180_all}(a) and (b), with parameters summarized in Table~\ref{tab:ex_cfg}. The top thick metal is used as the pixelated design layer, and the process stackup follows the foundry specifications. After one-time extraction of \(\mathbf{Z}_{\mathrm{ALL}}\), two representative random patterns with \(K=8\) edge I/O ports are evaluated in Fig.~\ref{fig:CMOS180_all}(c) and (d). MAPES agrees closely with CST over 30--100\,GHz, confirming its accuracy for RFIC-scale pixelated layouts.}
	
	\subsection{CMOS 65 nm Double-Layer Example}
	

	
\textcolor{black}{The CMOS 65\,nm double-layer configuration with vias is shown in Fig.~\ref{fig:CMOS65_all}(a) and (b), with parameters summarized in Table~\ref{tab:ex_cfg}. The structure uses the top thick metal and Metal~8, with inter-layer vias and a continuous ground plane in the outer ring of the top thick metal layer. After extracting \(\mathbf{Z}_{\mathrm{ALL}}\), two representative dual-layer patterns with \(K=8\) edge I/O ports are evaluated in Fig.~\ref{fig:CMOS65_all}(c) and (d). MAPES agrees well with CST across 30--100\,GHz, with minor deviations mainly appearing near deep-null regions where numerical sensitivity is higher.}

\subsection{\textcolor{black}{MAPES-Assisted Optimization of Microwave Filters}}
	
\begin{table}[t]
	\centering
	\caption{\textcolor{black}{Optimized filter specifications and optimization cost using MAPES.}}
	\label{tab:filter_optimization_summary}
	\begingroup
	\color{black}
	\renewcommand{\arraystretch}{1.16}
	\setlength{\tabcolsep}{1.5pt}
	\scriptsize
	\begin{tabular}{
			>{\centering\arraybackslash}m{0.17\columnwidth}
			>{\centering\arraybackslash}m{0.19\columnwidth}
			>{\centering\arraybackslash}m{0.13\columnwidth}
			>{\centering\arraybackslash}m{0.14\columnwidth}
			>{\centering\arraybackslash}m{0.16\columnwidth}
			>{\centering\arraybackslash}m{0.13\columnwidth}
		}
		\hline
		\textbf{Filter} &
		\textbf{Passband} &
		\textbf{Freq. Points} &
		\textbf{Eval.} &
		\textbf{MAPES Opt. Time} &
		\textbf{CST Est.} \\
		\hline
		High-pass &
		5.0--6.0\,GHz &
		22 &
		2854 &
		10.0\,min &
		27.2\,day \\
		
		Low-pass &
		2.0--2.5\,GHz &
		23 &
		4868 &
		17.8\,min &
		46.4\,day \\
		
		Bandpass 1 &
		2.9--3.1\,GHz &
		21 &
		10582 &
		35.3\,min &
		100.9\,day \\
		
		Bandpass 2 &
		4.1--4.3\,GHz &
		21 &
		10893 &
		37.8\,min &
		103.9\,day \\
		\hline
	\end{tabular}
	
	\vspace{3pt}
	\begin{minipage}{0.94\columnwidth}
		\scriptsize
		The passband and stopband targets are \(S_{21}\ge -1.0\) dB and
		\(S_{21}\le -15.0\) dB, respectively.
	\end{minipage}
	\endgroup
\end{table}

\begin{figure}[t]
	\centering
	
	\begin{minipage}[t]{0.48\columnwidth}
		\centering
		\includegraphics[width=0.72\linewidth]{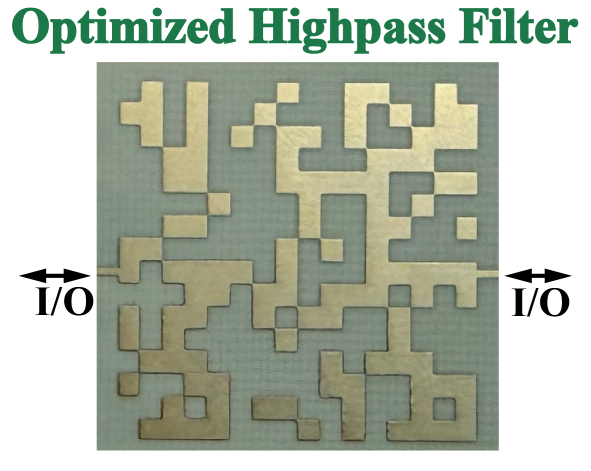}\\[-1pt]
		\includegraphics[width=0.82\linewidth]{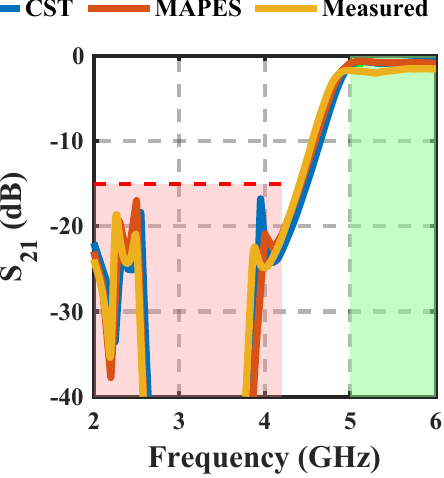}\\[-2pt]
		\textcolor{black}{\footnotesize (a)}
	\end{minipage}
	\hfill
	\begin{minipage}[t]{0.48\columnwidth}
		\centering
		\includegraphics[width=0.72\linewidth]{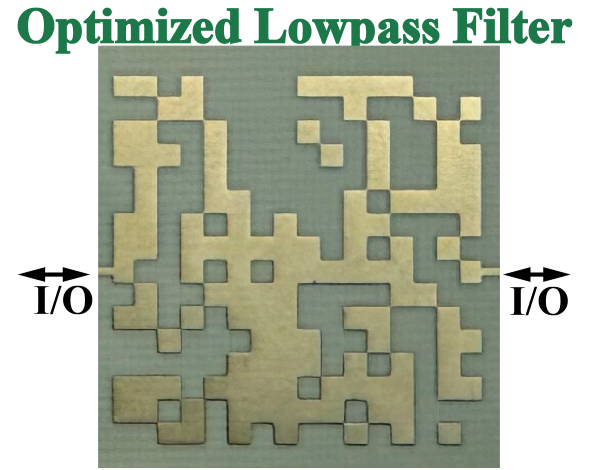}\\[-1pt]
		\includegraphics[width=0.82\linewidth]{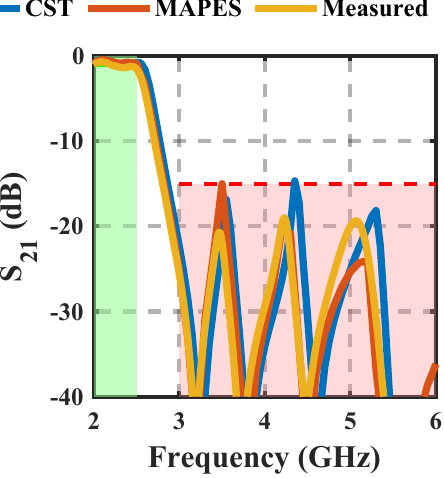}\\[-2pt]
		\textcolor{black}{\footnotesize (b)}
	\end{minipage}
	
	\vspace{5pt}
	
	\begin{minipage}[t]{0.48\columnwidth}
		\centering
		\includegraphics[width=0.72\linewidth]{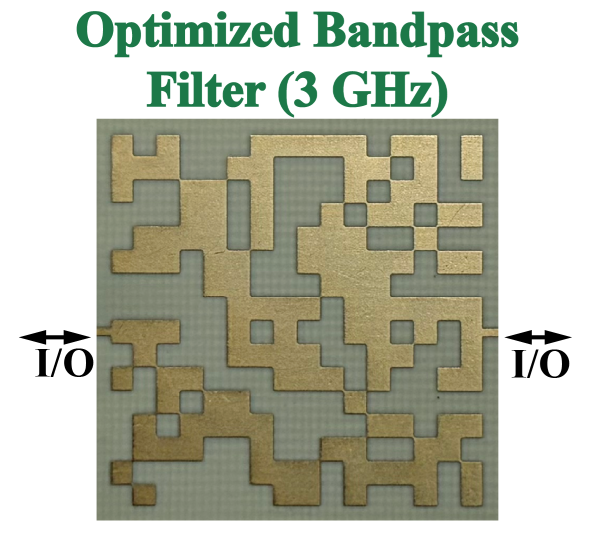}\\[-1pt]
		\includegraphics[width=0.82\linewidth]{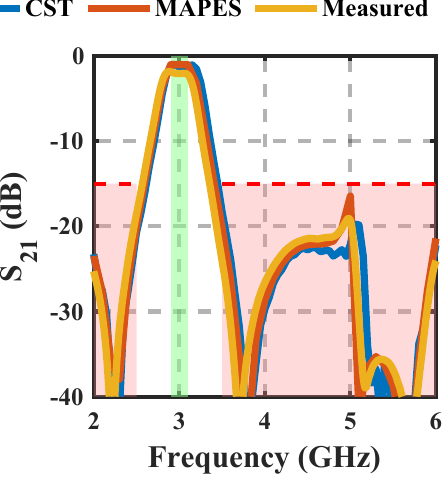}\\[-2pt]
		\textcolor{black}{\footnotesize (c)}
	\end{minipage}
	\hfill
	\begin{minipage}[t]{0.48\columnwidth}
		\centering
		\includegraphics[width=0.72\linewidth]{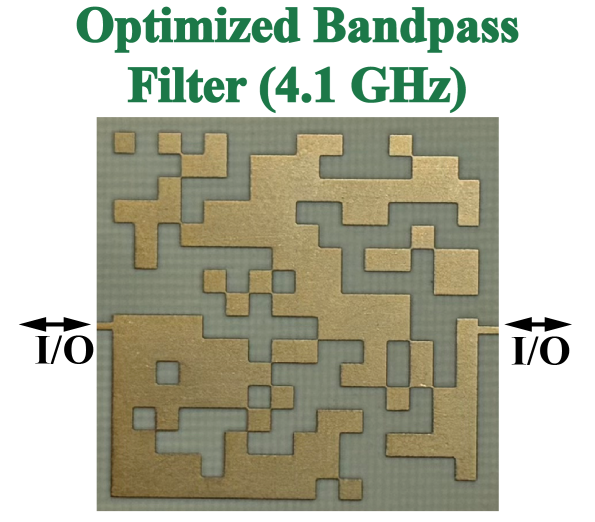}\\[-1pt]
		\includegraphics[width=0.82\linewidth]{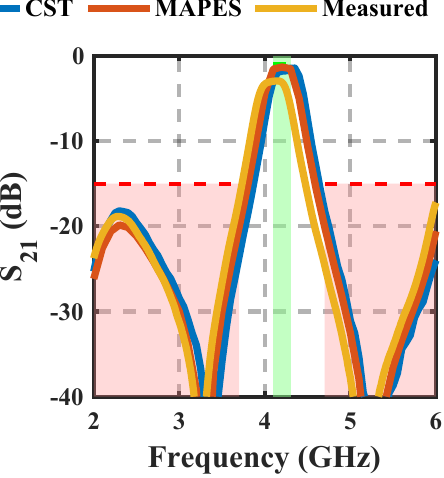}\\[-2pt]
		\textcolor{black}{\footnotesize (d)}
	\end{minipage}
	
	\caption{\textcolor{black}{MAPES-assisted optimization of practical pixelated microwave
			filters. The optimized pixel layouts and corresponding \(S_{21}\) responses are shown
			for four representative filters. (a) High-pass filter. (b) Low-pass filter. (c)
			Single-band bandpass filter around 3 GHz. (d) Single-band bandpass filter around
			4.1 GHz.}}
	\label{fig:optimized_filters}
\end{figure}

\begin{table*}[t]
	\caption{Comparison Between MAPES and CST for Different Configurations in Section~III}
	\label{tab:MAPES_vs_CST}
	\centering
	\renewcommand{\arraystretch}{1.2}
	\setlength{\tabcolsep}{1.5pt}
	
	\begin{tabular}{
			>{\centering\arraybackslash}m{0.145\textwidth}
			>{\centering\arraybackslash}m{0.13\textwidth}
			>{\centering\arraybackslash}m{0.055\textwidth}
			>{\centering\arraybackslash}m{0.09\textwidth}
			>{\centering\arraybackslash}m{0.09\textwidth}
			>{\centering\arraybackslash}m{0.095\textwidth}
			>{\centering\arraybackslash}m{0.09\textwidth}
			>{\centering\arraybackslash}m{0.09\textwidth}
			>{\centering\arraybackslash}m{0.085\textwidth}
		}
		\hline
		\textbf{Configuration (Section III)} &
		\textbf{Design Space} &
		\textbf{No. of I/O Ports} &
		\textbf{MAPES Time per Sim.} &
		\textcolor{black}{\textbf{MAPES Time per Freq. Point}} &
		\textbf{CST Time per Sim.} &
		\textbf{Mean Error ($E_{\mathrm{mean}}$)} &
		\textcolor{black}{\textbf{RMS Error ($E_{\mathrm{RMS}}$)}} &
		\textbf{Accel. Rate} \\ 
		\hline
		Single-Layer PCB & $16 \times 16$ (pixels) & 2 & 0.391\,s & \textcolor{black}{9.54\,ms} & 824\,s & 0.2199 & \textcolor{black}{0.2273} & $2107\times$ \\ 
		Double-Layer PCB & $16 \times 16 \times 3$ (pixels\&vias) & 2 & 1.609\,s & \textcolor{black}{39.24\,ms} & 991\,s & 0.1810 & \textcolor{black}{0.2801} & $615\times$  \\
		CMOS 180\,nm SL & $17 \times 17$ (pixels) & 8 & 1.028\,s & \textcolor{black}{14.48\,ms} & 720\,s & 0.0164 & \textcolor{black}{0.0269} & $700\times$ \\ 
		CMOS 65\,nm DL & $13 \times 13 \times 3$ (pixels\&vias) & 8 & 1.4138\,s & \textcolor{black}{19.91\,ms} & 2411\,s & 0.0275 & \textcolor{black}{0.0544} & $1705\times$ \\ 
		\hline
	\end{tabular}
	
	\vspace{2pt}
	\begin{minipage}{0.96\textwidth}
		\footnotesize
		\textbf{Note1:} All simulations were performed on a standard desktop computer equipped with an Intel i7-10700KF CPU and an NVIDIA GTX 3070 GPU. MAPES computations were executed in MATLAB R2024a on the same system. 
		
		\textcolor{black}{\textbf{Note2:} The reported MAPES time per simulation is the total MAPES evaluation time from an input pixel pattern to the output S-parameters, excluding the one-time extraction time of \(\mathbf{Z}_{\mathrm{ALL}}\). It includes the pixel-to-load mapping algorithm, matrix partitioning, matrix inversion, and parameter conversion. The \(\mathbf{Z}_{\mathrm{ALL}}\) extraction time is listed separately in Table~\ref{tab:ex_cfg}.}
	\end{minipage}
\end{table*}

\textcolor{black}{To further demonstrate that MAPES is not limited to analyzing randomly generated
	pixelated layouts, we used MAPES as the electromagnetic evaluator inside a pixel-level
	optimizer to synthesize practical microwave filter responses. The same single-layer
	PCB design domain in Section~III-A was used, namely a \(16\times16\) contiguous pixel
	array on Rogers RO4003C. The two I/O pixels were fixed at the left and right edges of the
	layout, and all other pixels were treated as binary optimization variables.}

\textcolor{black}{The optimization was performed using a sequential exhaustive binary optimization
	(SEBO) strategy \cite{Shen2016}. The design domain contains \(16\times16\) binary pixels. Starting from
	1000 random initial candidates, the pixels were randomly divided into groups of four, and
	all binary states within each group were exhaustively tested during each sweep, with a
	maximum of 50 sweeps. Each candidate layout was evaluated directly by MAPES using the
	precomputed \(\mathbf{Z}_{\mathrm{ALL}}\), without any CNN surrogate, retraining, or
	additional full-wave simulation. To reduce unnecessary cost, MAPES was evaluated only at
	selected frequency samples inside the prescribed passbands and stopbands, using 100-MHz
	passband sampling and 200-MHz stopband sampling, while transition-band points were excluded
	from the objective function. The objective penalizes insufficient passband transmission and
	insufficient stopband attenuation, with targets of \(S_{21}\ge -1\) dB in the passband and
	\(S_{21}\le -15\) dB in the stopband. The detailed filter specifications and optimization
	costs are listed in Table~\ref{tab:filter_optimization_summary}.}

\textcolor{black}{Fig.~\ref{fig:optimized_filters} shows the optimized layouts and the corresponding
	responses for four representative filters: a high-pass filter, a low-pass filter, and two
	single-band bandpass filters. In all cases, the optimized structures exhibit the desired
	filtering behavior. The MAPES predictions also agree well with CST and measured results,
	confirming that MAPES can guide the synthesis of useful microwave responses rather than
	only evaluate random pixel patterns.}

\textcolor{black}{The computational advantage of MAPES is especially clear in this iterative
	optimization task. As summarized in Table~\ref{tab:filter_optimization_summary}, the four
	optimizations required \(2854\)--\(10893\) candidate-layout evaluations over only
	\(21\)--\(23\) selected frequency points. Using the desktop computer in Table~\ref{tab:MAPES_vs_CST}, the total MAPES
	optimization time is only 10--38\,min. In contrast,
	evaluating the same number of candidate layouts by direct CST simulation would require
	approximately 27.2--103.9 days with the same desktop computer in Table~\ref{tab:MAPES_vs_CST}. Therefore, MAPES turns a multi-month
	full-wave optimization problem into a minute-scale optimization procedure while retaining
	close agreement with CST and measurement.}

\section{Comparison and Discussion}

\subsection{Accuracy and Efficiency vs. Full-Wave Simulations}

To quantitatively evaluate the accuracy and computational efficiency of the proposed MAPES framework, its predictions were systematically compared with full-wave electromagnetic simulations performed using CST Microwave Studio for all configurations presented in Section III. Table~\ref{tab:MAPES_vs_CST} summarizes the results of this comparison for each configuration. \textcolor{black}{Since MAPES can evaluate the response frequency point by frequency point, Table~\ref{tab:MAPES_vs_CST} also reports the average MAPES computation time per frequency point.}
The comparison metrics include mean magnitude error \(E_{\mathrm{mean}}\), \textcolor{black}{root-mean-square error \(E_{\mathrm{RMS}}\),} computational time, and efficiency improvement in terms of acceleration rate. 

\textcolor{black}{Here $E_{\mathrm{mean}}$ and $E_{\mathrm{RMS}}$ denote, respectively, the mean absolute error and the root-mean-square error of the magnitude deviation $\bigl|S_{ij}^{\mathrm{CST}}(f_m,n)-S_{ij}^{\mathrm{MAPES}}(f_m,n)\bigr|$, averaged over all I/O port pairs $(i,j)$, all sampled frequency points $f_m$, and all randomly generated examples $n$. Both metrics provide a comprehensive measure of the prediction accuracy of MAPES relative to CST and, as reported in Table~\ref{tab:MAPES_vs_CST}, lead to consistent conclusions.}

\begin{table*}[!t]
	\caption{Comparison Between MAPES and Other AI-Assisted EM Simulators}
	\label{tab:designspace_data}
	\centering
	\renewcommand{\arraystretch}{1.2}
	\setlength{\tabcolsep}{3pt}
	\begin{tabular}{
			>{\centering\arraybackslash}m{0.18\textwidth}
			>{\centering\arraybackslash}m{0.09\textwidth}
			>{\centering\arraybackslash}m{0.1\textwidth}
			>{\centering\arraybackslash}m{0.11\textwidth}
			>{\centering\arraybackslash}m{0.19\textwidth}
			>{\centering\arraybackslash}m{0.12\textwidth}
			>{\centering\arraybackslash}m{0.09\textwidth}
		}
		\hline
		\textbf{Configuration} & \textbf{Method} & \textbf{Design Space} & \textbf{Total No.\ of Possible Designs} & \textbf{Prior/Training Data Size$^{2}$ (EM Simulation Required)} & \textbf{Relative Percentage$^{1}$ (\%)} & \textbf{Potential Overfitting?}\\ 
		\hline
		Single-Layer PCB (Section.III A) & MAPES & $16 \!\times\! 16$ (pixels) & $2^{256}$ & $1444$ & $1.24 \times 10^{-72}$& NO  \\ \hline
		Double-Layer PCB (Section.III B) & MAPES & $16 \!\times\! 16 \!\times\! 3$ (pixels\&vias) & $2^{768}$ & $3144$ & $2.02 \times 10^{-226}$& NO \\ \hline
		Single-Layer CMOS 180\,nm (Section.III C) & MAPES & $17 \!\times\! 17$ (pixels) & $2^{289}$ & $1636$ & $1.64 \times 10^{-82}$& NO \\ \hline
		Double-Layer CMOS 65\,nm (Section.III D) & MAPES & $13 \!\times\! 13 \!\times\! 3$ (pixels\&vias) & $2^{507}$ & $2049$ & $4.80 \times 10^{-148}$& NO \\ \hline
		Single-Layer PCB \cite{Gupta2023princeton} & CNN & $12 \!\times\! 12$ (pixels) & $2^{142}$ & $500000$ & $8.96 \times 10^{-36}$& YES \\ \hline
		SiGe 90 nm \cite{Karahan2023princeton} & Deep-CNN & $16 \!\times\! 16$ (pixels) & $2^{256}$ & $340000$ & $2.93 \times 10^{-70}$& YES \\ \hline
		Single-Layer 90 nm SiGe BiCMOS \cite{Karahan2024princeton} & CNN & $25 \!\times\! 25$ (pixels) & $2^{625}$ & $386000$ & $2.77 \times 10^{-181}$& YES \\ \hline
		Single-Layer GF 22nm FDX+ \cite{Chu2025eth1} & Template-Based CNN & $16 \!\times\! 16 \!\times\! 2$ (pixels) & $2^{512}$ & $52546$ & $3.91 \times 10^{-146}$& YES \\ \hline
	\end{tabular}
	
	\vspace{3pt}
	\begin{minipage}{0.96\textwidth}
		\footnotesize
		\textbf{Note1:} The relative percentage represents the ratio between the prior-data/training data size and the entire combinational design space,  illustrating the extremely low sampling ratio required by the proposed MAPES framework for effective prior-data simulations.\\
		\textbf{Note2:} The prior-data simulations for MAPES were performed using the CST time-domain solver, which requires exciting the virtual ports sequentially. Consequently, the number of full-wave simulation runs equals the number of virtual ports. In contrast, a frequency-domain solver in CST or HFSS can simulate all virtual ports within a single run to extract the complete multiport matrix, potentially leading to a further reduction in total pre-simulation time for MAPES.
	\end{minipage}
\end{table*}

The single-layer and double-layer PCB configurations exhibit mean errors of 0.2199 and 0.1810, respectively. These relatively larger deviations are mainly attributed to slight frequency shifts in the resonant response, as observed in Figs.~\ref{fig:PCB1_all}(c),(d) and~\ref{fig:PCB2_all}(c),(d). Nevertheless, the overall amplitude agreement between MAPES and CST remains high across the entire frequency range, indicating that the physical accuracy of MAPES is well maintained. In practical microwave and RF circuit design, these levels of deviation are entirely acceptable, since most design optimization processes can readily account for small frequency offsets. \textcolor{black}{The physical origin of these PCB-level deviations, together with a simple scaling calibration that largely removes them, is analyzed in Section~IV-C in conjunction with the selection of the model parameters.}

In contrast, the two integrated-circuit configurations (the CMOS 180\,nm single-layer and CMOS 65\,nm double-layer designs) demonstrate substantially lower mean errors of 0.0164 and 0.0275, respectively. The higher operating frequency and finer metal patterning in IC layouts reduce the impact of boundary discretization compared to PCB-scale structures. The agreement observed in Figs.~\ref{fig:CMOS180_all}(c),(d) and~\ref{fig:CMOS65_all}(c),(d) further confirms that MAPES can accurately simulate RFIC layouts while being extremely efficient.

Regarding computational efficiency, MAPES demonstrates dramatic acceleration compared with CST simulations. For example, in the single-layer PCB case, MAPES required only \(0.391\,\mathrm{s}\) per simulation, whereas CST required approximately \(824\,\mathrm{s}\), yielding an acceleration factor exceeding \(2100\times\); the other three configurations achieve acceleration rates from \(615\times\) to \(1705\times\). \textcolor{black}{This advantage stems from the analytical formulation in~\eqref{eq:7}, which obtains the entire EM response via a single matrix operation once \(\mathbf{Z}_{\mathrm{ALL}}\) is established. Moreover, since the I/O-port number only affects the dimension of the extracted submatrix rather than requiring additional field solutions, the MAPES per-evaluation time stays nearly constant as the port count grows, whereas the CST time increases approximately linearly. This makes MAPES particularly attractive for complex multiport MW/RFIC designs.}

\textcolor{black}{For a fair accounting of the total cost, the one-time extraction time of \(\mathbf{Z}_{\mathrm{ALL}}\) is also listed in Table~\ref{tab:ex_cfg}. These prior simulations were performed on the standalone server described in Table~\ref{tab:ex_cfg}. Even after including this one-time cost, MAPES remains substantially faster when many pixel patterns are evaluated. For example, for the CMOS 65\,nm double-layer case, \(10^6\) design evaluations take about 17\,d\,14\,h\,52\,min with MAPES on the standard desktop computer used for the per-pattern evaluations, not on the server used for \(\mathbf{Z}_{\mathrm{ALL}}\) extraction. This runtime could be significantly shortened with server-level computing resources. By contrast, direct CST simulation would require approximately 76.5 years on the same desktop computer.}

\subsection{Comparison With Other AI-Assisted EM Emulators}

Table~\ref{tab:designspace_data} summarizes a detailed comparison between the proposed MAPES and several state-of-the-art AI-assisted EM simulators reported in recent literature, highlighting two key advantages.

\textcolor{black}{First, for the same or even larger design spaces, MAPES requires only a small fraction of the EM simulations needed by AI emulators. For a $16\times16$ single-layer array, MAPES uses only $1444$ prior simulations, roughly $1\%$ of the hundreds of thousands of samples typically required by CNN-/deep-CNN-based emulators~\cite{Gupta2023princeton,Karahan2023princeton,Karahan2024princeton}. This prior dataset can be generated within a few days on a mid-range workstation, instead of weeks of high-performance parallel computing. Similar conclusions hold for the double-layer PCB and RFIC examples, where MAPES uses only a few thousand prior simulations, corresponding to less than $10^{-148}\%$ of the total design space.}

\textcolor{black}{Second, MAPES provides reliable predictions over the entire design space because it follows directly from first-principles multiport network theory: $\mathbf{Z}_{\mathrm{ALL}}$ encapsulates the coupling among all pixels, and any new layout is evaluated in closed form via~\eqref{eq:7}, guaranteeing physical consistency and eliminating statistical uncertainty. Data-driven emulators, in contrast, are trained on a vanishingly small subset of the design space and may degrade on unseen or irregular pixel patterns. The smaller training set reported in~\cite{Chu2025eth1} relies on a template-seeded strategy specific to transformer geometries and is thus not generalizable to other topologies such as filters or couplers.}

\textcolor{black}{The main cost of MAPES is the matrix inversion in~\eqref{eq:7}, whose size grows with the number of virtual ports. Consequently, although MAPES is far faster than full-wave solvers, it can be slower than a purely AI-based surrogate that evaluates a layout through a single GPU forward pass; e.g.,~\cite{Karahan2023princeton} reports $4096$ evaluations in about $0.3$\,s on a high-end server. Part of this gap reflects hardware differences, but it also reflects a fundamental difference between a physics-driven evaluator and a trained surrogate.}

\subsection{Modeling Assumptions and Scaling Parameter Selection}

\textcolor{black}{MAPES replaces the original contiguous pixel pattern with an equivalent virtual-pixel network. This approximation relies on two assumptions: (i) the coupling between physically disconnected virtual pixels is weak, and (ii) replacing the original pixels with virtual pixels to insert virtual ports does not significantly change the EM response. These assumptions are accurate when each pixel is electrically small, but they can introduce a small residual frequency shift when the pixels become electrically larger.}

\textcolor{black}{This explains the higher prediction error observed in the PCB examples. In the single-layer and double-layer PCB cases, each pixel measures approximately $0.016\lambda$ and $0.008\lambda$ at the center frequency of 4\,GHz. Compared with the original contiguous pixel pattern, the equivalent virtual-pixel model contains virtual-port gaps and corner paths, which slightly increase the effective electrical path length. As a result, the MAPES response shows a small downward frequency shift relative to the CST reference. In contrast, the CMOS pixels are electrically much smaller, approximately $0.004\lambda$, so this path-length effect is negligible.}

\textcolor{black}{The geometry of the equivalent model can be controlled by two parameters: the virtual-pixel size ratio $\beta$ and the global scaling ratio $\alpha$:
	\begin{equation}
		W_{\mathrm{vp}} = \alpha W_{\mathrm{orig}}\beta,\qquad
		W_{\mathrm{gap}} = \alpha W_{\mathrm{orig}}(1-\beta),
		\label{eq:alpha_beta}
	\end{equation}
	where $W_{\mathrm{orig}}$ is the original pixel width. The parameter $\beta$ controls the relative size of the virtual pixels and gaps, while $\alpha$ uniformly scales the entire equivalent model to compensate for residual electrical path-length error.}

\begin{figure}[t]
	\centering
	\begin{minipage}{0.24\textwidth}
		\centering
		\includegraphics[width=1\linewidth]{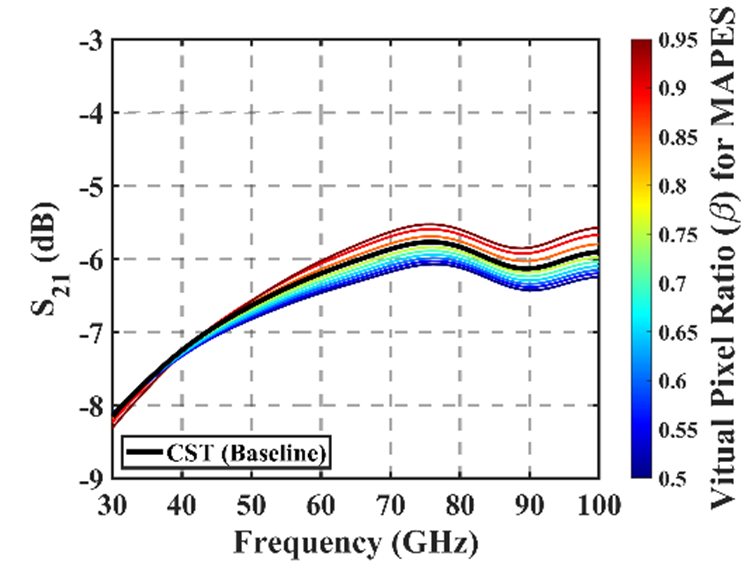}\\
		(a)
	\end{minipage}
	\hfill
	\begin{minipage}{0.24\textwidth}
		\centering
		\includegraphics[width=1\linewidth]{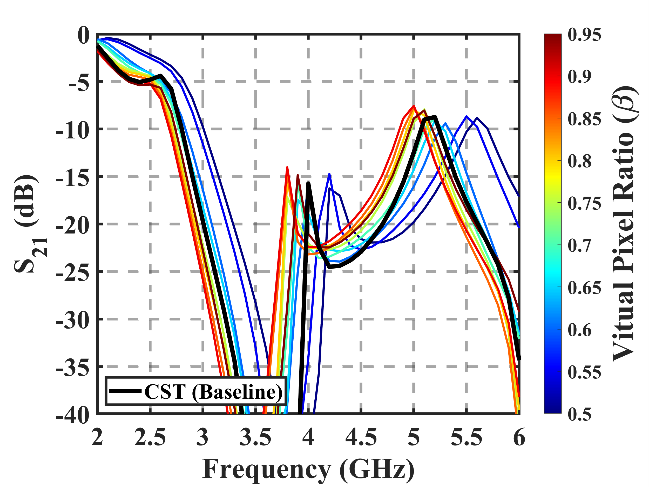}\\
		(b)
	\end{minipage}
	\caption{\textcolor{black}{Sensitivity analysis of $\beta$. MAPES-predicted S-parameters for $\beta$ from 0.50 to 0.95 (coloured curves) compared with CST full-wave simulation (black): (a) the CMOS 180\,nm single-layer configuration (30--100\,GHz, $\varepsilon_r\approx 11.9$) and (b) the PCB single-layer configuration (2--6\,GHz, $\varepsilon_r\approx 3.55$)}}
	\label{fig:beta_cmos180_and_pcb}
\end{figure}

\textcolor{black}{The choice of $\beta$ involves a trade-off. A large $\beta$ leaves very small gaps and can introduce residual coupling between disconnected pixels, whereas a small $\beta$ overly distorts the original geometry. To quantify this effect, $\beta$ was swept from $0.50$ to $0.95$ with $\alpha=1.0$. As shown in Fig.~\ref{fig:beta_cmos180_and_pcb}, both the CMOS 180\,nm and PCB single-layer cases show stable predictions for $\beta\approx0.70$--$0.85$, while $\beta>0.90$ causes excessive residual coupling and $\beta<0.65$ causes geometric distortion. The four examples in Section~III used $\beta=0.833$, $0.819$, $0.842$, and $0.869$, respectively, confirming that this range is robust across different technologies and frequencies.}

\begin{figure}[t]
	\centering
	\begin{minipage}{0.24\textwidth}
		\centering
		\includegraphics[width=1.1\linewidth]{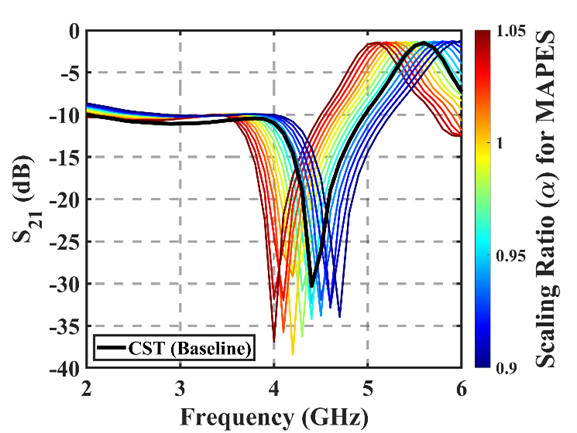}\\
		(a)
	\end{minipage}
	\hfill
	\begin{minipage}{0.24\textwidth}
		\centering
		\includegraphics[width=0.9\linewidth]{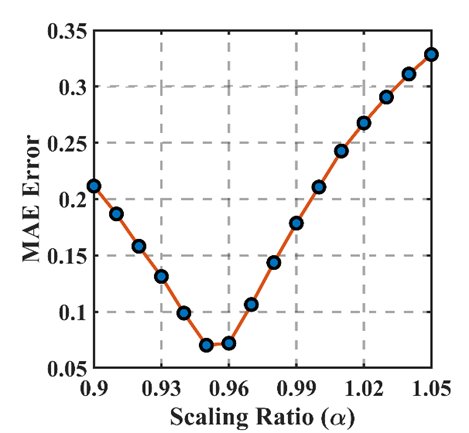}\\
		(b)
	\end{minipage}
	\caption{\textcolor{black}{First dual-layer pixel pattern (PCB double-layer) in Fig. \ref{fig:PCB2_all}: (a) MAPES-predicted $S_{21}$ for different scaling ratios $\alpha$ compared with the CST baseline, and (b) MAE as a function of $\alpha$, showing a clear minimum near $\alpha\approx0.96$.}}
	\label{fig:scaling_alpha_ex1}
	
	\centering
	\begin{minipage}{0.24\textwidth}
		\centering
		\includegraphics[width=1.1\linewidth]{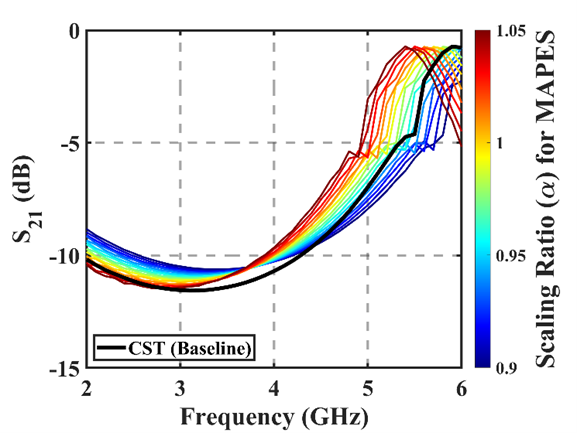}\\
		(a)
	\end{minipage}
	\hfill
	\begin{minipage}{0.24\textwidth}
		\centering
		\includegraphics[width=0.9\linewidth]{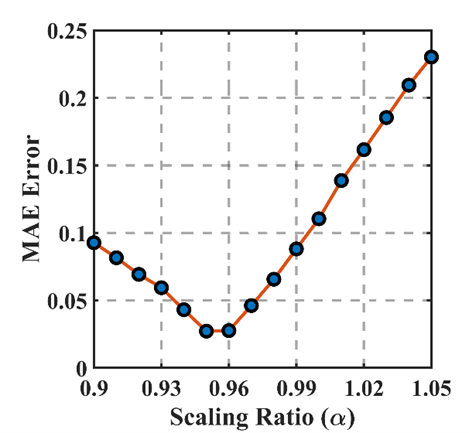}\\
		(b)
	\end{minipage}
	\caption{\textcolor{black}{Second dual-layer pixel pattern (PCB double-layer) in Fig. \ref{fig:PCB2_all}: (a) MAPES-predicted $S_{21}$ for different scaling ratios $\alpha$ compared with the CST baseline, and (b) MAE as a function of $\alpha$, showing a clear minimum near $\alpha\approx0.96$.}}
	\label{fig:scaling_alpha_ex2}
\end{figure}

\textcolor{black}{The global scaling ratio $\alpha$ is used only to correct residual frequency shift. This correction is mainly useful for PCB-scale layouts, where the pixels are electrically larger. By choosing $\alpha$ slightly smaller than unity, the total virtual-pixel unit dimension is reduced, which shortens the effective signal path and compensates for the extra electrical length introduced by the virtual-port gaps. For RFIC-scale layouts and electrically small pixels, $\alpha=1$ is sufficient. For the double-layer PCB examples in Figs.~\ref{fig:scaling_alpha_ex1} and~\ref{fig:scaling_alpha_ex2}, $\alpha\approx0.96$ reduces the MAE by over $60\%$, showing that the PCB-level errors are not fundamental limitations of MAPES but can be compensated by a simple scaling calibration.}

\textcolor{black}{For practical use, $\beta$ should be selected in the stable range around $0.70$--$0.85$, while $\alpha$ can be set to 1 for RFIC-scale layouts and slightly below 1 for PCB-scale layouts when a residual frequency shift is observed. Detailed selection guidelines are provided in Appendix~B.}

\subsection{Comparison with Precomputation-Based Efficient Calculators}

\begin{table}[t]
	\textcolor{black}{\caption{Comparison of MAPES With Precomputation-Based Methods}
		\label{tab:mapes_vs_mom_pngf}
		\centering
		\renewcommand{\arraystretch}{1.25}
		\setlength{\tabcolsep}{2pt}
		\footnotesize
		\newcommand{\yes}{\textcolor{green!55!black}{\checkmark}}
		\newcommand{\no}{\textcolor{black}{\boldmath$\times$}}
		\begin{tabular}{
				>{\raggedright\arraybackslash}m{0.20\columnwidth}
				>{\centering\arraybackslash}m{0.20\columnwidth}
				>{\centering\arraybackslash}m{0.22\columnwidth}
				>{\centering\arraybackslash}m{0.24\columnwidth}
			}
			\hline
			\textbf{Aspect} & \textbf{MoM~\cite{Johnson1999}} & \textbf{PNGF~\cite{Sun2025}} & \textbf{MAPES (This Work)} \\
			\hline
			Precomp.\ operator & Z-matrix (mesh-level) & Green function $G$ (Yee-cell level) & \textbf{$\mathbf{Z}_{\mathrm{ALL}}$ (virtual-port level)} \\
			Core matrix dim.$^{\dagger}$ & $\sim\!18N_{\mathrm{pixel}}$ & $\sim\!24N_{\mathrm{pixel}}$ & \textbf{$\sim\!6N_{\mathrm{pixel}}$} \\
			Relative inversion cost$^{\dagger}$ & $27\times$ & $64\times$ & \textbf{$1\times$ (baseline)} \\
			Decoupled from mesh density & \no\ No & \no\ No & \textbf{\yes\ Yes} \\
			Commercial-solver workflow & \no\ Custom code & \no\ Custom FDTD/FDFD & \textbf{\yes\ Standard multiport sim.} \\
			Multi-layer \& vias & \no\ Not native & \no\ Not native & \textbf{\yes\ Native} \\
			Arbitrary I/O ports & \no\ Restricted & \no\ Restricted & \textbf{\yes\ Yes} \\
			Process/stackup generality & \no\ Limited & \no\ Limited & \textbf{\yes\ High} \\
			\hline
		\end{tabular}
		\vspace{2pt}\\
		\begin{minipage}{0.96\columnwidth}
			\scriptsize $^{\dagger}$Assuming $3\!\times\!3$ mesh cells per pixel for MoM and PNGF; the relative inversion cost follows the cubic scaling of dense matrix inversion, i.e., $(18/6)^3\!=\!27$ for MoM and $(24/6)^3\!=\!64$ for PNGF. Finer meshes further widen the gap.
	\end{minipage}}
\end{table}

\begin{figure}[t]
	\centering
	\begin{minipage}{0.47\columnwidth}
		\centering
		\includegraphics[width=\linewidth]{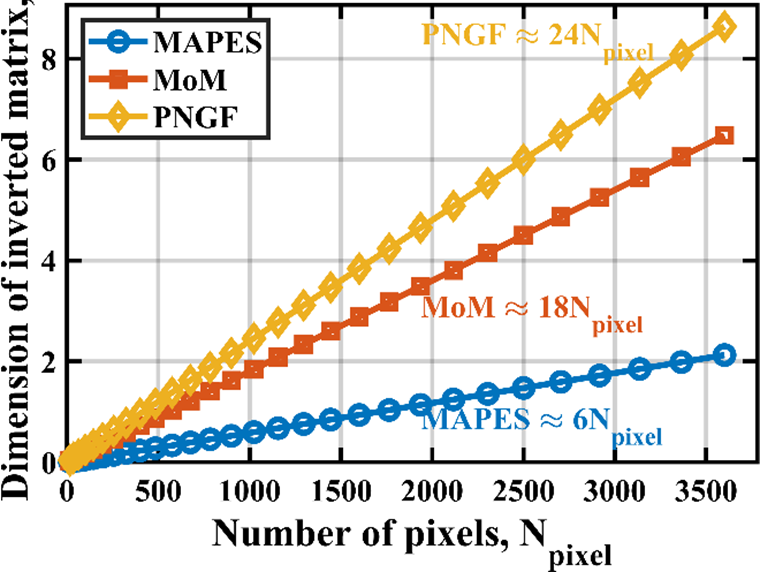}\\
		(a)
	\end{minipage}\hfill
	\begin{minipage}{0.49\columnwidth}
		\centering
		\includegraphics[width=\linewidth]{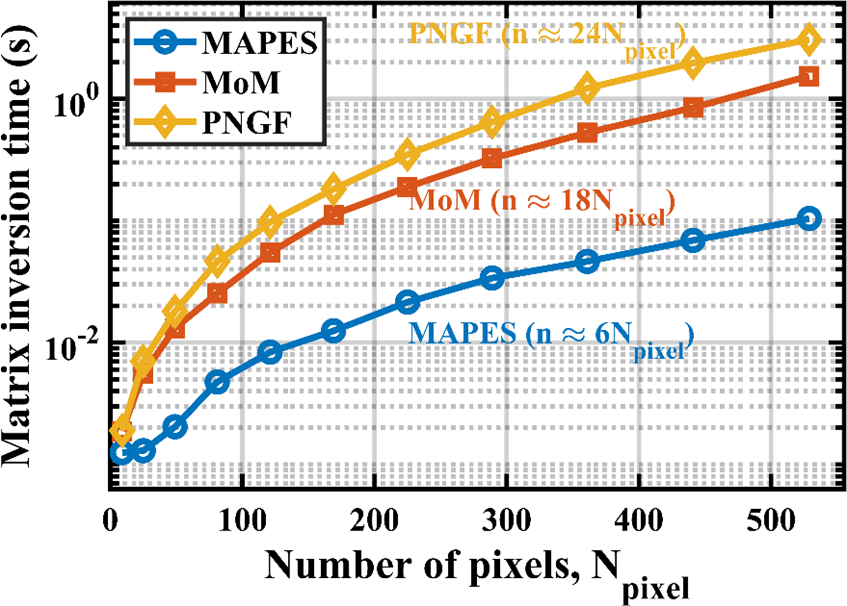}\\
		(b)
	\end{minipage}
	\caption{\textcolor{black}{Matrix-size and timing comparison of MAPES with MoM- and PNGF-sized matrices under $3\!\times\!3$ mesh-per-pixel settings. (a)~Representative scaling of the inverted matrix dimension with the number of pixels ($N_{\mathrm{pixel}}$); each pixel uses 18 triangular elements in MoM and 24 in-plane Yee-cell unknowns in PNGF, versus about 6 virtual ports in MAPES. (b)~Measured per-evaluation matrix inversion time versus $N_{\mathrm{pixel}}$ for single-layer configurations. All computations were performed on a standard desktop computer equipped with an Intel i7-10700KF CPU and an NVIDIA GTX 3070 GPU using MATLAB R2024a.}}
	\label{fig:matrix_scaling_time}
\end{figure}

\textcolor{black}{Beyond AI-assisted surrogates, two representative precomputation-based approaches have been proposed to accelerate pixel-level electromagnetic evaluation. In the GA/MoM method~\cite{Johnson1999}, a ``mother'' impedance matrix is filled once for a fully metallized structure using the Method of Moments; candidate substructures are then evaluated by extracting the corresponding submatrix, thereby avoiding repeated matrix-fill operations. More recently, the PNGF method~\cite{Sun2025} precomputes a numerical Green function that encapsulates the static portions of the simulation environment into a single matrix; any candidate design is evaluated by solving a reduced linear system whose dimension equals only the number of unknowns inside the optimization region.}

\textcolor{black}{While MAPES shares the high-level philosophy of precomputing an electromagnetic operator and reusing it across layout variations, its \emph{port-level} formulation sets it apart from both the mother-structure MoM approach~\cite{Johnson1999} and the PNGF framework~\cite{Sun2025}. As summarized in Table~\ref{tab:mapes_vs_mom_pngf}, the key distinctions are: (i)~a fundamentally smaller core matrix ($\sim\!6N_{\mathrm{pixel}}$ virtual ports versus $\sim\!18N_{\mathrm{pixel}}$ for MoM and $\sim\!24N_{\mathrm{pixel}}$ for PNGF, giving roughly $27\times$ and $64\times$ lower per-evaluation inversion cost under the same coarse mesh); (ii)~decoupling of the core matrix size from the full-wave mesh density, so refining the mesh to improve $\mathbf{Z}_{\mathrm{ALL}}$ does not enlarge the MAPES model; and (iii)~native compatibility with commercial-solver workflows, since $\mathbf{Z}_{\mathrm{ALL}}$ is obtained from standard multiport simulations without access to internal mesh matrices or custom FDTD/FDFD code. These advantages, together with native support for multi-layer stacks, vias, and arbitrary I/O ports, distinguish MAPES from both prior methods. A detailed discussion of the three distinctions is provided in Appendix~C.}

\textcolor{black}{\subsection{Scalability and Physical Properties}}

\textcolor{black}{The dominant per-evaluation cost in MAPES is the inversion of the $(Q-K)\times(Q-K)$ matrix $(\mathbf{Z}_{L}+\mathbf{Z}_{VP,VP})$ in~\eqref{eq:7}. For a single-layer $M\times N$ array, $Q\approx 6MN$ from~\eqref{eq:1}, giving a cost of $\mathcal{O}((6MN)^3)$. \textcolor{black}{Fig.~\ref{fig:matrix_scaling_time}(b) reports the measured MAPES inversion time against MoM- and PNGF-sized matrices on the workstation used throughout this study, while Fig.~\ref{fig:matrix_scaling_time}(a) shows the corresponding growth of the inverted matrix dimension.} For all four configurations in Section~III ($Q$ up to $\approx 3{,}100$), the inversion time stays well below one second; even for a hypothetical $50\!\times\!50$ array ($Q=14{,}704$) it is only on the order of seconds, still orders of magnitude faster than a full-wave CST simulation.}

\textcolor{black}{In terms of storage, $\mathbf{Z}_{\mathrm{ALL}}$ requires $Q^2\times 8$ bytes per frequency point, ranging from $\approx16$\,MB ($Q=1{,}444$) to $\approx75$\,MB ($Q=3{,}144$) for the demonstrated cases, and $\approx1.6$\,GB for a $50\!\times\!50$ array, all accommodable on standard workstations. Because $\mathbf{Z}_{\mathrm{ALL}}$ varies smoothly with frequency, interpolation or model-order reduction can further reduce the required frequency samples. Several established techniques are directly applicable for future acceleration, including Woodbury low-rank updates~\cite{Bayin2018} when few pixels change between iterations, exploitation of the diagonal structure of $\mathbf{Z}_{L}$, and GPU-accelerated inversion.}

\textcolor{black}{The reduced model also preserves the key physical properties by construction. Since $\mathbf{Z}_{\mathrm{ALL}}$ is extracted from a linear, reciprocal environment ($\mathbf{Z}_{\mathrm{ALL}}=\mathbf{Z}_{\mathrm{ALL}}^{\mathrm{T}}$) and $\mathbf{Z}_{L}$ is a real diagonal (hence symmetric) matrix, the Schur complement in~\eqref{eq:7} preserves symmetry, so $\mathbf{Z}_{\mathrm{MAPES}}$ remains \emph{reciprocal} for any pixel pattern. As $\mathbf{Z}_{\mathrm{ALL}}$ is positive-real (passive)~\cite{Pozar2021} and $\mathbf{Z}_{L}$ contains only non-negative real impedances, the Schur complement is also positive-real, guaranteeing \emph{passivity} under all activation patterns. Finally, across all four configurations (matrix sizes up to $\approx3{,}000$) the condition number of the inverted matrix stays on the order of $10^5$, well within the stable range of double-precision arithmetic; for much larger arrays, standard regularization or pivoted LU factorization can be employed.}

\section{Conclusion}
This work proposed and validated the MAPES, a physics-driven analytical electromagnetic simulator for arbitrary pixel-based MW circuits and RFIC. By introducing diagonal virtual pixels and systematically placed virtual ports, MAPES captures horizontal/vertical/diagonal coupling in a compact multiport impedance matrix extracted from only a limited number of full-wave simulations, enabling closed-form prediction of any configuration via multiport network theory. Across single- and multi-layer PCB and CMOS examples, MAPES matches full-wave CST S-parameters while achieving \(600\text{--}2000\times\) speedup; unlike data-driven AI simulators, it avoids overfitting, generalizes across the entire design space, and naturally supports vias and multilayer interconnects. \textcolor{black}{The framework is further shown to preserve reciprocity, passivity, and numerical stability by construction, ensuring physically consistent predictions across all configurations.} MAPES can also serve as a fast dataset generator or a physics backbone for hybrid physics-informed neural networks, and future work will explore GPU-accelerated and AI-assisted matrix inversion to further improve speed.

\appendices

\textcolor{black}{\section{Extraction Procedure of $\mathbf{Z}_{\mathrm{ALL}}$}}

\textcolor{black}{The prior matrix $\mathbf{Z}_{\mathrm{ALL}}$ is extracted by a standard
multiport simulation in CST/HFSS through four steps. (1)~\textit{Build the model}: draw the
virtual pixels and diagonal virtual pixels of Section~II.A on the actual substrate stackup
(same dielectric, metal thickness, conductivity, and ground plane as the target design).
(2)~\textit{Place ports}: assign a discrete port at every virtual-port location and
orientation, i.e., between adjacent virtual pixels (horizontal/vertical), between diagonal
virtual pixels and their neighbors (diagonal), between outer virtual pixels and ground
(I/O candidates), and between overlapping pixels on adjacent layers (vias), giving $Q$ ports
in total as in~\eqref{eq:1} and~\eqref{eq:2}. (3)~\textit{Set reference impedance}: assign all
ports a common 50\,$\Omega$ reference; the value is immaterial because the matrix is
renormalization-invariant. (4)~\textit{Simulate and convert}: run one multiport simulation
over the band to obtain the $Q\times Q$ scattering matrix and convert it to
$\mathbf{Z}_{\mathrm{ALL}}$. A frequency-domain solver excites all ports in one run; a
time-domain solver excites them sequentially ($Q$ runs). The matrix is computed once and
reused for all pixel patterns via~\eqref{eq:7}, using only routine discrete-port S-parameter
simulation with no access to internal mesh matrices or custom code.}

\textcolor{black}{\section{Practical Selection of $\alpha$ and $\beta$}}

\textcolor{black}{The analysis in Section~IV-C gives the following practical guidelines:
	\begin{itemize}
		\item \emph{Virtual pixel size ratio $\beta$:} choose $\beta\approx0.70$--$0.85$ (about $0.85$ in this work). The prediction accuracy is insensitive within this range, while extreme values ($\beta\!>\!0.90$ or $\beta\!<\!0.65$) should be avoided.
		\item \emph{Global scaling ratio $\alpha$:} set $\alpha=1$ for RFIC processes and electrically small PCB pixels; use a slightly smaller $\alpha$ (such as 0.96) for PCB-scale structures when a residual frequency shift is observed.
		\item \emph{Optional fine calibration of $\alpha$:} if higher accuracy is required, $\alpha$ can be refined using a few random sample patterns by comparing the full-wave response of the actual pixel layout with that of its equivalent virtual-pixel structure under different $\alpha$ values.
\end{itemize}}

\textcolor{black}{\section{Detailed Distinctions From Precomputation-Based Methods}}

\begin{figure}[t]
\centering
\includegraphics[width=1\columnwidth]{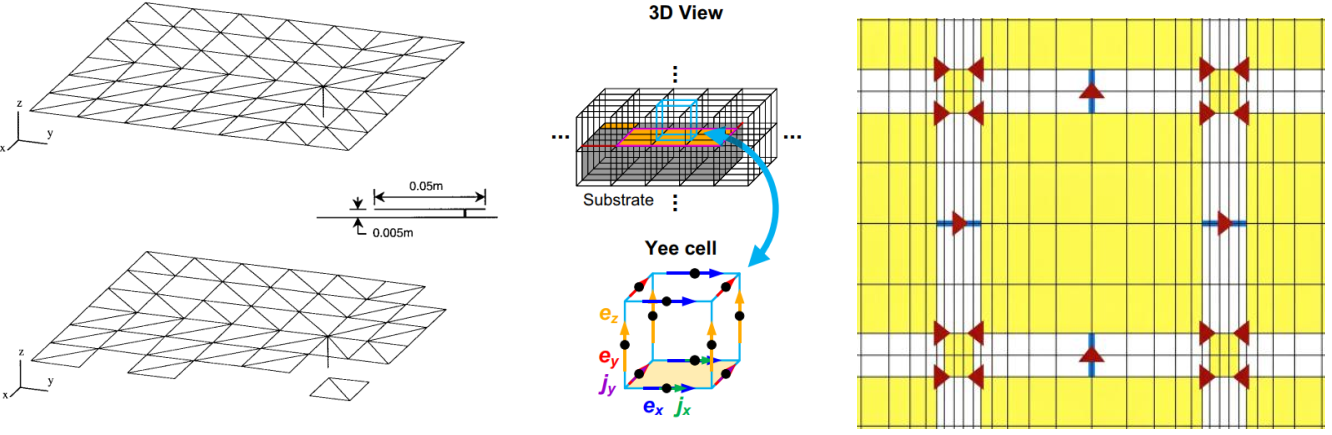}
\\
(a) \hspace{2.5cm} (b) \hspace{2.5cm} (c)
\caption{\textcolor{black}{Comparison of the modelling level in representative precomputation-based electromagnetic approaches. (a) Triangular RWG-basis discretization in the mother-structure MoM method~\cite{Johnson1999}. (b) Yee-cell field/current discretization in the PNGF method~\cite{Sun2025}. (c) MAPES equivalent model with virtual pixels, diagonal virtual pixels, and virtual ports.}}
\label{fig:modelling_level}
\end{figure}

\textcolor{black}{This appendix elaborates the three distinctions summarized in Table~\ref{tab:mapes_vs_mom_pngf} and Figs.~\ref{fig:modelling_level} and~\ref{fig:matrix_scaling_time}.}

\textcolor{black}{\textit{Distinction 1: Port-level formulation with fundamentally smaller core matrices.} All precomputation-based methods ultimately evaluate performance by inverting a core matrix, whose dimension is the critical difference (Fig.~\ref{fig:matrix_scaling_time}(a)). In MoM~\cite{Johnson1999}, each pixel requires about 18 RWG basis functions (for a $3\!\times\!3$ rectangular sub-meshing), giving $\sim\!18 N_{\mathrm{pixel}}$. In PNGF~\cite{Sun2025}, each pixel tile comprises about 24 Yee-cell unknowns, giving $\sim\!24 N_{\mathrm{pixel}}$. In MAPES, only about 6 virtual ports per pixel are needed, giving $\sim\!6 N_{\mathrm{pixel}}$. Since inversion cost scales cubically, MAPES achieves a $27\times$ reduction over MoM $((18/6)^3=27)$ and a $64\times$ reduction over PNGF $((24/6)^3=64)$ under the same coarse mesh; finer meshes further widen the gap.}

\textcolor{black}{\textit{Distinction 2: Decoupling from full-wave mesh density.} In MAPES, the core matrix dimension is set solely by the number of virtual ports, which scales with the number of pixels, not with mesh density. Refining the mesh to improve $\mathbf{Z}_{\mathrm{ALL}}$ does not change the analytical model size or the per-evaluation cost. In both MoM~\cite{Johnson1999} and PNGF~\cite{Sun2025}, any mesh refinement proportionally enlarges the core matrix, coupling computational cost to accuracy requirements.}

\textcolor{black}{\textit{Distinction 3: Native compatibility with commercial solver workflows.} MAPES obtains $\mathbf{Z}_{\mathrm{ALL}}$ from standard multiport S-/Z-parameter simulations in commercial tools (CST, HFSS), a routine operation requiring no access to internal mesh matrices or custom code. In MoM~\cite{Johnson1999}, the mesh impedance matrix is generally inaccessible in commercial software and requires specialized coding; PNGF~\cite{Sun2025} can partially leverage commercial tools but still requires custom FDTD/FDFD implementations for full functionality.}

\bibliographystyle{IEEEtran}
\bibliography{ref}

\end{document}